\documentclass{easychair}

\usepackage[T1]{fontenc}
\usepackage{setspace}
\usepackage{verbatim}
\usepackage{upquote}  
\usepackage{amssymb}
\usepackage{graphicx}
\usepackage{color}
\usepackage{setspace}
\usepackage{multicol}
\usepackage{array}
\usepackage{enumitem}
\usepackage{hyperref}
\usepackage{fontawesome5}  

\newcommand{\smalltt}[1]{\small \texttt{#1}}
\newcommand{\example}{{\scriptsize \faFile[regular]}}
\newcommand{\tildenot}{\raisebox{0.4ex}{\texttildelow}}
\newenvironment{packed_itemize}{
\vspace*{-0.3em}
\begin{itemize}
\setlength{\partopsep}{0pt}
\setlength{\itemsep}{1pt}
\setlength{\parskip}{0pt}
\setlength{\parsep}{0pt}
}{\end{itemize}}
\newenvironment{packed_enumerate}{
\vspace*{-0.3em}
\begin{enumerate}
\setlength{\partopsep}{0pt}
\setlength{\itemsep}{1pt}
\setlength{\parskip}{0pt}
\setlength{\parsep}{0pt}
}{\end{enumerate}}
\AddToHook{cmd/ttfamily/after}{\frenchspacing}
\title{The TPTP Format for Interpretations}

\author{
  Geoff Sutcliffe\inst{1}
\and
  Alexander Steen\inst{2}
\and
  Pascal Fontaine\inst{3}
\and
  Lydia Kondylidou\inst{4}
}

\institute{
  University of Miami,
  Miami, USA\\
  \email{geoff@cs.miami.edu}
\and
  University of Greifswald,
  Greifswald, Germany\\
  \email{alexander.steen@uni-greifswald.de}
\and
  University of Li{\`e}ge,
  Li{\`e}ge, Belgium\\
  \email{Pascal.Fontaine@uliege.be}
\and
  Ludwig-Maximilians-Universit{\"a}t München
  Munich, Germany\\
  \email{Lydia.Kondylidou@tcs.ifi.lmu.de}
}

\authorrunning{Sutcliffe, Steen, Fontaine, Kondylidou}
\titlerunning{TPTP World Interpretations}

\begin{document}

\maketitle
\begin{abstract}
This paper describes the TPTP format for representing interpretations.
It provides a background survey that helped ensure that the representation format is adequate
for different types of interpretations: Tarskian, Herbrand, and Kripke interpretations.
The needs of applications that use models are considered.
The syntax and semantics of the format are expounded in detail, with multiple examples.
Verification of models is discussed.
Some tools that support processing the format are noted.
The properties of interpretations represented in the format are discussed.
\end{abstract}
\section{Introduction}
\label{Introduction}

Historically, Automated Theorem Proving (ATP) has, as the name suggests, focused largely on the
task of proving theorems from axioms -- the derivation of conclusions that follow inevitably 
from known facts~\cite{RV01-HAR}.
The axioms and the conjecture to be proved (and hence become a theorem) are written in an 
appropriately expressive logic, and the proofs are often similarly written in logic~\cite{SS+06}.
The converse task of disproving conjectures, proving that a conjecture is not a theorem of the 
axioms, is another facet of interest, e.g.,~\cite{Win82,LM87,McC98,Zha05,BN10-ITP,RKK17,EPS24}.
This process depends on finding a \emph{countermodel} for the conjecture, i.e., finding an 
\emph{interpretation} (a structure that assigns meaning to the symbols of the language of the
formulae, and consequently maps formulae to truth values) that is a \emph{model} of the axioms 
(maps the axioms to \textit{true}) but not a model for the conjecture (maps the conjecture to 
\textit{false}).
A salient application area that uses this form of ATP is verification~\cite{DKW08}, where a 
countermodel is used to pinpoint the reason why a proof obligation fails, and correspondingly 
points to the location of the fault in the system being verified.
Other applications of model finding include checking the consistency of an axiomatization 
\cite{SS+17}, operations research~\cite{Hoo93}, commonsense reasoning~\cite{KR94}, and solving 
model finding problems~\cite{Win82}.

The TPTP World~\cite{Sut17,Sut24} is the well established infrastructure that supports research, 
development, and deployment of 
ATP systems.\footnote{%
\href{https://www.tptp.org}{\tt www.tptp.org}}
Various parts of the TPTP World have been deployed in a range of applications, in both academia 
and industry.
The TPTP World includes the TPTP problem library~\cite{Sut09}, 
the TSTP solution library~\cite{Sut10}, 
tools and services for processing ATP problems and solutions~\cite{Sut10}, 
and it supports the CADE ATP System Competition (CASC)~\cite{Sut16}.
The TPTP language~\cite{Sut23-IGPL} is one of the keys to the success of the TPTP World.
Originally the TPTP language supported only first-order clause normal form (CNF)~\cite{SS98-JAR}.
Over the years full first-order form (FOF)~\cite{Sut09}, 
typed first-order form (TFF)~\cite{SS+12,BP13-TFF1}, 
typed extended first-order form (TXF)~\cite{SK18}, 
typed higher-order form (THF)~\cite{SB10,KSR16}, 
dependently typed higher-order form (DHF)~\cite{RK+25},
and non-classical forms (NXF, NHF)~\cite{SF+22}
have been added.
The TPTP language is used for writing ATP problems, derivations, and, most relevant to this 
work, \emph{interpretations}~\cite{SS+06,Sut08-KEAPPA}.

A TPTP format for interpretations with finite domains has previously been defined~\cite{SS+06},
and has served the ATP community adequately for almost 20 years. 
The old format is output by several \emph{model finder} ATP systems, e.g.,~Paradox~\cite{CS03}, 
FM-Darwin~\cite{BF+06}, Vampire~\cite{KV13}.
The need for a format for interpretations with infinite domains, and for a format for Kripke 
interpretations~\cite{Kri63}, has led to the development of the (new current) TPTP format for 
interpretations.
This work describes the format.
The underlying principle is unchanged: interpretations are represented in formulae written in
the TPTP language.

\paragraph{Related work:}
There are other concrete representations of interpretations in use:
The SMT-LIB standard~\cite{BFT17} defines a format for model output, and commands to inspect 
models.  
SAT solvers generally output models as specified by the SAT competitions~\cite{JL+12}, in a 
simple format similar to the DIMACS input format~\cite{Bab93}.
Some individual model finders have defined their own formats for models, e.g.,~the 
output formats of Nitpick~\cite{BN10-ITP} and Z3~\cite{dMB08}.

\paragraph{This paper is organized as follows:}
Section~\ref{Interpretations} discusses the nature of interpretations, considering what is
needed from interpretations, and the various forms that interpretations can take.
Section~\ref{NewTarskian} defines the format for Tarskian interpretations, and
Section~\ref{NewKripke} does the same for Kripke interpretations.
Section~\ref{Conclusion} concludes and discusses plans for future work.

\paragraph{Note well:} 
This paper refers to examples of problems and their interpretations that are listed in the 
Appendices.
They are referenced with a \example{} icon.
 
\section{About Interpretations}
\label{Interpretations}

\subsection{What do we Need?}
\label{Need}

The needs of applications that use model finding vary according to their use of the model.
In the simplest case, applications need to know only that a model exists.
Examples of such applications include checking the consistency of an axiomatization~\cite{CI15},
use as a subroutine in more complex reasoning, e.g.,~for
axiom selection~\cite{SP07,Pud07-ESARLT}, and establishing the existence of a bug in a
verification process\footnote{%
Bill McCune claimed that establishing the existence of a bug without having an explicit model
to help pinpoint the bug would be ``frustrating''. And he should have known.}.
A key weakness of a model finder that claims to have found a model but does not output an 
explicit model is that it is necessary to trust the model finder.

In many applications it is necessary to have an explicit model, in some representation format that
allows for analysis of the model.
Applications that productively use an explicit model include finding inconsistencies in 
axiomatizations~\cite{SS+17}, identification of bugs in verification~\cite{CE82,QS82},
studying modal and description logics~\cite{SH13}, solving problems encoded as a model finding 
problems~\cite{Win82}, and evaluating formulae wrt the model~\cite{SS+23-LPAR}.
Manual inspection of explicit models can also be useful, e.g.,~\cite{EK+10}.
A key strength of a model finder that outputs an explicit model is that the model can be verified,
i.e., it is not necessary to trust the model finder.

Given the innate desirability of obtaining explicit models of satisfiable formulae, the desirable
properties of interpretation representation are as follows:
\begin{packed_enumerate}
\item Interpretations must be for the symbols of the input formulae.
\item Evaluation of any formulae wrt an interpretation should be possible, because many uses of
      interpretations can be reduced to evaluation~\cite[\S3.2]{CLP04}.
      The evaluation should be tractable, or in a tractable core.
      This corresponds to the \emph{atom test} and \emph{formula evaluation} postulates suggested
      by~\cite{FL96,CLP04}, and might be considered to be the most important property of
      interpretation representation.
\item Verification of models should be possible by checking that the problem formulae evaluate 
      to \textit{true} wrt the model.
      The verification should be tractable, or in a tractable core.
\item Interpretations should be sufficiently (human) comprehensible to be useful in (human)
      applications.
\end{packed_enumerate}

\subsection{What do we Have?}
\label{Have}

A \emph{Tarskian interpretation}~\cite{TV56} consists of a non-empty domain of distinct elements 
(not the Herbrand Universe -- see Herbrand interpretations below) for each type used in the 
formulae (just one domain for untyped logic), mappings for function symbols from tuples of 
domain elements to the domain, and mappings for predicate symbols from tuples of domain elements 
to $\{true,false\}$~\cite{Hun96,Gal15}.
For higher-order logic only Henkin interpretations~\cite{Hen50} are considered.
An overview of some ways of building and representing Tarskian interpretations is provided 
by~\cite{CLP04}, and~\cite{Pel03-EQMC} provides a comprehensive list of approaches.

Two types of Tarskian interpretations are clear (and more might exist):
\begin{packed_enumerate}
\item \emph{Finite} Tarskian interpretations have only finite domains.
      The domain and symbol mappings can be explicitly enumerated.
      Finite models are typically produced by starting with domains of size one, and incrementing 
      the sizes until a model is found.
      At each iteration the formulae are reduced to a decidable logic, e.g.,~
      propositional~\cite{CS03,McC03-MACE4-TR} or function free~\cite{BF+09} logic, and an ATP 
      system for that logic is used to decide if there is a model.
      There are several ATP systems that produce finite Tarskian models, e.g.,~Paradox, FM-Darwin, 
      and Vampire.
\item \emph{Infinite} Tarskian interpretations have one or more infinite domains.
      Infinite domains can be explicitly generated, e.g.,~terms representing Peano numbers, or 
      implicitly specified, e.g.,~some set of algebraic numbers such as the integers~\cite{BB13}.
      There are some ATP systems that produce infinite Tarskian models, e.g.,~
      cvc5~\cite{BB+22-cvc5}, FEST~\cite{EPS24}, and Z3.
\end{packed_enumerate}

A \emph{Herbrand interpretation}~\cite{Her30} is a particular form of Tarskian interpretation.
A Herbrand interpretation has the Herbrand universe as the domain, the mapping for function symbols
is the ``identity'', and the mapping for predicate symbols is from the Herbrand base to 
$\{true,false\}$.
Every set of formulae induces its set of Herbrand models.
Formulae in Skolem normal form induce Herbrand models with a Herbrand universe that includes any 
Skolem symbols. 
An empty set of formulae induces all interpretations, i.e., all interpretations are models
of the set.
Such sets of formulae can be the result of applying model preserving transformations to the 
problem formulae (even the input itself induces its set of Herbrand models), or be generated 
by an ATP system with the explicit intention of representing Herbrand models.
They are important in the context of resolution and all similar modern calculi; saturations (see 
below) are a common case.
\begin{packed_itemize}
\item Formulae that are intended to represent Herbrand interpretations are written 
      \emph{HI-formulae} in this paper, to avoid confusion.
      Examples include saturations (discussed separately in the next bullet point), 
      Bachmair-Ganzinger models~\cite{BG94,Lyn13}, a disjunction of implicit generalisations 
      (DIGs)~\cite{LM87}, sets of constrained unit clauses~\cite{CZ92,CP95,CP95-TAB}, SGGS clause 
      sequences~\cite{BP16}, tableaux~\cite{Hah01}
      and hyper-tableaux~\cite{BFN96}, eq-interpretations~\cite{Pel03-EQMC}, and 
      predicate definitions over the term algebra~\cite{SK12}.
      E-KRHyper~\cite{PW07} was an ATP system that produced hyper-tableaux, and
      iProver~\cite{Kor08,SK12} is an ATP system that outputs predicate definitions 
      over the term algebra.
\item \emph{Saturations}~\cite{BG+01,Pel03-JSC} are a special case of HI-formulae.
      A saturation is a fixed point for a set of clauses at which further application of a
      complete inference system does not generate any new clauses (up to redundancy).
      A saturation of a set of clauses determines the same set of Herbrand models as 
      the clauses themselves.
      Saturations can be used as explicit models in equational theories~\cite{JRS26-arXiv}.
      Saturation-based ATP systems include E~\cite{SCV19}, Prover9~\cite{McC-Prover9-URL}, 
      Vampire, and Zipperposition~\cite{VB+21}.
\end{packed_itemize}
While the domain of a Herbrand interpretation determined by Herbrand formulae is known to be the 
Herbrand universe, there might not be an explicit symbol interpretation that can be used 
constructively by users.

A \emph{Kripke interpretation}~\cite{Kri63} adds a layer of \emph{possible worlds} over Tarskian 
interpretations.
There can be a finite or infinite number of worlds.
There is an \emph{accessibility relation} between the worlds, which can be subject to the
requirements of the logic being used, e.g.,~for modal logic \textbf{M} the accessibility 
relation must be reflexive~\cite{Gar18}.
Within each world there is a Tarskian interpretation. 
The Tarskian interpretations' domains might be \emph{constant}, i.e., the same in all the worlds,
but might also vary.
The designation of symbols might be \emph{rigid}, i.e., the same in all the worlds' Tarskian 
interpretations, but might also vary.
The designation of ground terms might be \emph{local}, i.e., from only the domain of the world,
or might be global.
Formulae can be evaluated wrt Kripke interpretations with a finite worlds and local terms by
evaluating modal operators using Kripke semantics, and
evaluating formulae within a world wrt the Tarskian interpretation in the world.

\subsection{What does the TPTP format Provide?}
\label{TPTPProvides}

The TPTP format represents interpretations in \emph{interpretation-formulae} that have the new
annotated formula role \smalltt{interpretation}.
The details are provided in Sections~\ref{NewTarskian}, \ref{NewHerbrand} and~\ref{NewKripke}.
Interpretation-formulae are written using the TPTP syntax.
As a brief reminder of the TPTP syntax, problems and solutions are built from 
{\em annotated formulae} of the form: \\
\hspace*{0.5cm}{\em language}{\tt (}{\em name}{\tt ,}
{\em role}{\tt ,}
{\em formula}{\tt ,}
{\em source}{\tt ,}
{\em useful\_info}{\tt )} \\
The {\em language}s supported are \smalltt{cnf} (clause normal form), \smalltt{fof}
(first-order form), \smalltt{tff} (typed first-order form), and \smalltt{thf}
(typed higher-order form).
The {\em role}, e.g., \smalltt{axiom}, \smalltt{lemma}, \smalltt{conjecture}, defines the 
use of the formula.
In a {\em formula}, terms and atoms follow Prolog conventions -- functions and predicates start 
with a lowercase letter or are {\tt '}single quoted{\tt '}, and variables start with an uppercase 
letter.
The language also supports interpreted symbols that either start with a {\tt \$}, e.g., the 
truth constants \smalltt{\$true}, \smalltt{\$false}, the individual and boolean
types \smalltt{\$i}, \smalltt{\$o}, or are composed of 
non-alphabetic characters, e.g., integer/rational/real numbers such as 27, 43/92, -99.66.
The logical connectives in the TPTP language are
{\tt !>}, {\tt ?*}, {\tt @+}, {\tt @-}, {\tt !}, {\tt ?}, {\tt {\tildenot}}, {\tt |}, {\tt \&}, 
{\tt =>}, {\tt <=}, {\tt <=>}, and {\tt <{\tildenot}>}, for the mathematical connectives
$\Pi$, $\Sigma$, choice (indefinite description), definite description,
$\forall$, $\exists$, $\neg$, $\vee$, $\wedge$, $\Rightarrow$, $\Leftarrow$, $\Leftrightarrow$, 
and $\oplus$ respectively.
Equality and inequality are expressed as the infix operators {\tt =} and {\tt !=}.
The {\em source} and {\em useful\_info} are optional.

The notions of interpretations, models, partial interpretations, finite interpretations,
Herbrand interpretations, etc., are captured in the SZS 
ontologies~\cite{Sut08-KEAPPA}.\footnote{%
Updated at
\href{https://tptp.org/UserDocs/SZSOntology/}{\tt tptp.org/UserDocs/SZSOntology}}
For this work the Success ontology was extended with a new value ModelExtending (MEX).
The success ontology provides values for a pair Ax-C, where Ax is a set (conjunction) of axioms 
and C is a conjecture to be proved from Ax. 
MEX is defined as ``Some interpretations are models of Ax, and some interpretations are models 
of C, and all models of C are conservative extensions of models of Ax (which means that all 
models of C are models of Ax)''.
This is used for transformations on satisfiable sets of formulae in which the models of the set 
are unchanged, or are changed only by adding new domain elements or mappings, so that the
extended models are still models of the original formulae. 
Examples of such transformations are Skolemization, and adding logical consequences of a set 
into the set.

\subsection{Do we Have what we Need?}
\label{HaveNeed}

\begin{figure}[htbp]
\centering
\includegraphics[width=0.75\textwidth]{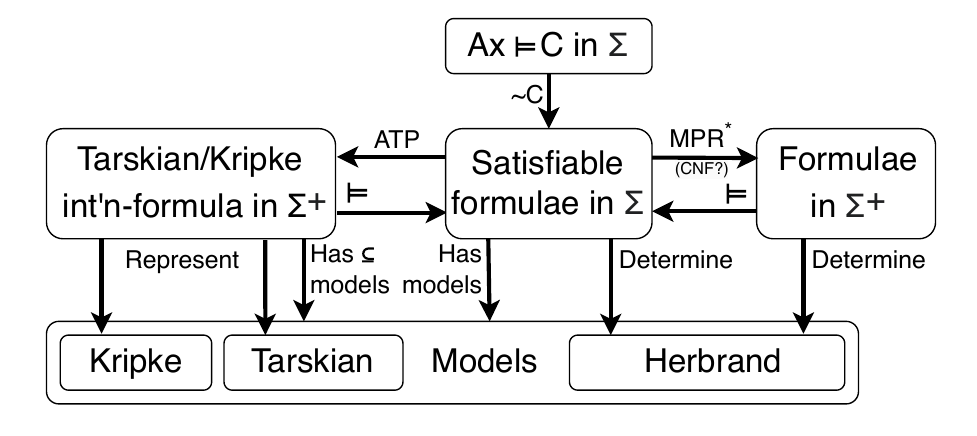}
\caption{A Landscape of Model Building}
\label{ModelLandscape}
\end{figure}

Figure~\ref{ModelLandscape} provides an overview of the situation.
The starting point is the set of {\sf Satisfiable formulae} written in the language $\Sigma$. 
The satisfiable formulae might have been formed from 
{\sf Ax $\cup$ \{\tildenot{}C\}}. 
Going down leads to the {\sf Models} of the satisfiable formulae.
Going left from the {\sf Satisfiable formulae in $\Sigma$} is the pathway taken by ATP systems 
that find a Tarskian/Kripke model, and output interpretation-formulae representing the model.
The interpretation-formulae can be for a conservative extension $\Sigma^+$ of the language of 
the input formula, e.g.,~by the addition of Skolem symbols, defined symbols, etc.
If the interpretation-formulae correctly represent a model of the satisfiable formulae, then
the satisfiable formulae can be proved ($\vDash$) from the interpretation-formulae.
The models of the interpretation-formulae are conservative extensions of a subset (by the 
definition of $\vDash$) of the models of the satisfiable formulae.
Going right from the {\sf Satisfiable formulae in $\Sigma$} is the pathway taken by ATP systems
(for classical logics at least) that apply 
transformations to the 
satisfiable set, to produce formulae that determine Herbrand models of the satisfiable formulae.
The satisfiable formulae can be proved ($\vDash$) from the sets that result from applying the 
transformations.

\subsubsection{Formula Evaluation}
\label{FormulaEvaluation}

The second, most important, property of interpretation representation is to evaluate formulae
wrt the interpretation.
With interpretation-formulae this can be achieved as follows:
\begin{itemize}
\item For Tarskian interpretation-formulae with finite domains, direct evaluation can be used.
      \begin{packed_itemize}
      \item Interpret quantifiers using Tarskian semantics.
      \item Interpret ground (grounded with domain elements) terms and atoms using the mappings.
      \item Interpret formulae using the truth tables for connectives.
      \end{packed_itemize}
      This approach is taken internally in Vampire to verify its finite models.
\item For Tarskian interpretation-formulae, theorem proving can be used.
      \begin{packed_itemize}
      \item If a formula can be proved from the interpretation-formulae then it is \textit{true} 
            in the interpretation represented by the interpretation-formulae.
      \item If a formula can be disproved from the interpretation-formulae then it is 
            \textit{false} in the interpretation represented by the interpretation-formulae.
      \item In practice, a formula might be neither proved nor disproved within reasonable 
            resource limits, so that nothing is known.
      \end{packed_itemize}
\item For Herbrand interpretations determined by HI-formulae, theorem proving can be 
      used:~\footnote{%
      \emph{We (well, at least the first author) believe that this verification technique works,
      and so do Stephan Schulz and Uwe Waldmann, but Andrei Voronkov and Christoph Weidenbach say 
      they do not.}}
      \begin{packed_itemize}
      \item If a formula can be proved from the HI-formulae then it is \textit{true} in all 
            the Herbrand interpretations determined by the HI-formulae.
      \item If a formula can be disproved from the HI-formulae then it is \textit{false} in 
            some of the Herbrand interpretations determined by the HI-formulae.
      \item In practice, a formula might be neither proved nor disproved within reasonable 
            resource limits, so that nothing is known.
      \end{packed_itemize}
\item For Kripke interpretation-formulae, theorem proving can be used.
      This is described in Section~\ref{KripkeVerification}.
\end{itemize}

\subsubsection{Model Verification}
\label{Verification}

The third property of interpretation representation is to verify a model of given formulae.
Given ways to evaluate a formula wrt interpretation-formulae, models represented in
interpretation-formulae can be checked.
This has (at least) the following aspects: 
\begin{packed_enumerate}
\item Verify that the type declarations and interpretation-formulae are syntactically well-formed 
      and well-typed. 
\item Validate that the interpretation-formulae correctly represent the model found by the AT
      system.
\item Verify that the interpretation-formulae are satisfiable.
\item Verify that the interpretation represented by the interpretation-formulae is a model for 
      the given formulae.
\end{packed_enumerate}

These steps can be (in)completed as follows:
\begin{enumerate}
\item This can be confirmed using standard parsing and type checking tools, 
      e.g.,~\cite{VS06,HR15,Ste26}.
\item This cannot be confirmed without insight into the ATP system that found the model.
\item This can be confirmed using a trusted model finder.
      Hopefully, confirming that the interpretation-formulae are satisfiable is much easier than 
      finding the model itself, so the ATP system used to check the satisfiability can be weaker 
      and more trusted than the ATP system that found the model.
      In light of the next point, this model finding task can be made easier by combining the
      interpretation-formulae with the problem formulae (axioms and negated conjecture).
\item This can be confirmed using the theorem proving approach to verification, in which the 
      problem formulae $\Phi$ are proved from the interpretation-formulae $\varphi$ using a trusted 
      theorem prover.
      The soundness of this approach is proved by showing that if $\Phi$ can be proved from 
      $\varphi$ then the interpretation $I$ represented by $\varphi$ is a model of $\Phi$.
      This is done for finite Tarskian interpretations for FOF in~\cite{SS+23-LPAR}.
      We believe the proof in~\cite{SS+23-LPAR} lifts naturally to TFF and THF, but is technically 
      more complicated due to the introduction of types.
      The extension to infinite domains is quite simple after that.
      For Kripke interpretation-formulae it is necessary to make changes that reduce the
      verification problem to TXF/THF (see Section~\ref{KripkeVerification}).
\end{enumerate}

Steps~2 to 4 are implemented in the AGMV model verifier~\cite{SS+23-LPAR}, available in the
SystemOnTSTP~\cite{Sut07-CSR} web interface.\footnote{%
\href{https://www.tptp.org/cgi-bin/SystemOnTSTP}{\tt www.tptp.org/cgi-bin/SystemOnTSTP}}
      
\section{The Format for Tarskian Interpretations}
\label{NewTarskian}

A simple Tarskian interpretation-formula is a conjunction of components
(see the examples [\example{}:~\ref{FOF_Finite.s}, \ref{TFF_Finite.s}, 
\ref{TFF_Finite_SeparateDomains.s}, \ref{THF_Finite.s}]):
\begin{packed_itemize}
\item A domain for each type in the problem formulae.
\item Interpretation of non-boolean symbols, as equalities whose left-hand sides are formed from 
      symbols applied to domain elements, and whose right-hand sides are domain elements.
\item Interpretation of boolean symbols, as literals formed from symbols applied to domain 
      elements; positive literals are \emph{true} and negative literals are \emph{false}.
\end{packed_itemize}
The TPTP format for Tarskian interpretation-formulae requires no changes to the TPTP
language syntax (other than the addition of the \smalltt{interpretation} role that is used for
interpretation-formulae).
At this stage only monomorphic types are supported, but the syntax is flexible enough to extend
the format to polymorphic types.

Function and predicate symbols are interpreted by applying them directly to domain elements 
(\emph{{\`a} la}~\cite[\S5.3.4]{Gal15}), rather than using the presentation of interpretations that 
introduces a new interpretation function for each function/predicate symbol~\cite[\S5.3.2]{Gal15}.
This approach obviates the need for introducing interpretation functions, and makes 
interpretation of formulae using theorem proving simpler.
However, applying the problem's function and predicate symbols to domain elements that are 
defined with different types to the problem's types requires using \emph{type-promotion} 
bijections to keep interpretation-formulae well-typed (see Section~\ref{TarskianNewFiniteTypes}).

[\example{}:~\ref{FOF_Finite.s}, \ref{TFF_Finite.s}, \ref{TFF_Peano.s}, \ref{TFF_Integer.s}, 
\ref{THF_Finite.s}] illustrate the components of finite Tarskian interpretations in a single 
interpretation-formula, as explained in Sections~\ref{FOFTarskian}, \ref{TarskianReusingTypes}, 
and \ref{TarskianNewFiniteTypes}.
[\example{}:~\ref{FOF_Finite_Medium.s}, \ref{FOF_Finite_Fine.s}, \ref{TFF_Finite_Medium.s}, 
\ref{TFF_Finite_Fine.s}, \ref{THF_Finite_Medium.s}] illustrate Tarskian interpretations split 
over multiple interpretation-formulae, as explained in Section~\ref{NewTarskianSplit}.
[\example{}:~\ref{TFF_Finite_Compact.s}] illustrates a Tarskian interpretation that is compacted 
using quantification, as explained in Section~\ref{NewTarskianCompact}.
[\example{}:~\ref{FOF_Formulae.s}, \ref{FOF_Saturation.s}] illustrate interpretation-formulae 
that determine Herbrand interpretations, as explained in Section~\ref{NewHerbrand}.
These examples illustrate the possibilities for writing Tarskian interpretations in the TPTP 
format, but are not intended to be exhaustive.
In particular, interpretations of higher-order formulae have yet to be fleshed out.

\subsection{Finite Interpretations without Types}
\label{FOFTarskian}

[\example{}:~\ref{FOF_Finite.p}] shows a FOF problem that has a countermodel with a finite 
domain, which is shown in [\example{}:~\ref{FOF_Finite.s}].

The domain specification of a finite interpretation without types is a conjunction of:
\begin{packed_itemize}
\item Specification of the domain elements as a universally quantified disjunction of equalities 
      whose right-hand sides are the domain elements, 
      e.g.,~[\example{}:~\ref{FOF_Finite.s}]:\\
      \hspace*{0.5cm}\smalltt{! [X] : ( X = "a" | X = "f" | X = "john" | X = "gotA")}
\item Specification of the distinctness of the domain elements.
      Distinctness can be specified with pairwise inequalities, with the \smalltt{\$distinct} 
      predicate, or by using \smalltt{"double quoted"} terms as domain elements (different 
      \smalltt{"double quoted"} terms are known to be unequal (and often not used in problems,
      which makes them easy to identify as domain elements)),
      e.g.,~[\example{}:~\ref{FOF_Finite.s}] \smalltt{"a"} is known to be unequal to \smalltt{"f"}.
\end{packed_itemize}

The function mappings are equalities between domain elements and functions applied to domain 
elements,
e.g.,~[\example{}:~\ref{FOF_Finite.s}]: \\
\hspace*{0.5cm}\smalltt{grade\_of("john") = "f"}\\
The predicate mappings are literals, e.g.,~[\example{}:~\ref{FOF_Finite.s}]: \\
\hspace*{0.5cm}\smalltt{created\_equal("john","gotA")}\\
If \smalltt{"double quoted"} constants are already used in the formulae being interpreted, they
are naturally interpreted as themselves.
Note that for a complete interpretation all function and predicate mappings need to be provided,
even if they make no sense in a typed sense, e.g.,~[\example{}:~\ref{FOF_Finite.s}]: \\
\hspace*{0.5cm}\smalltt{grade\_of("f") = "a"}\\
\hspace*{0.5cm}\smalltt{\tildenot{}\,created\_equal("a","john")}

[\example{}:~\ref{FOF_Finite.s.p}] shows the verification problem for the model.

\subsection{Finite Interpretations reusing Problem Types}
\label{TarskianReusingTypes}

[\example{}:~\ref{TFF_Finite.p}] shows a TFF problem that has a countermodel with finite domains,
which is shown in [\example{}:~\ref{TFF_Finite.s}].

Each domain specification is a conjunction of:
\begin{packed_itemize}
\item Specification of the domain elements as a universally quantified disjunction of equalities 
      whose right-hand sides are the domain elements, e.g.,~[\example{}:~\ref{TFF_Finite.s}]:\\
      \hspace*{0.5cm}\smalltt{! [DC: cat]: ( DC = d\_garfield | DC = d\_arlene | DC = d\_nermal )}
\item Specification of the distinctness of the domain elements, 
e.g.,~[\example{}:~\ref{TFF_Finite.s}]:\\
      \hspace*{0.5cm}\smalltt{\$distinct(d\_garfield,d\_arlene,d\_nermal)}
\end{packed_itemize}

The function mappings are equalities between domain elements and functions applied to domain 
elements, e.g.,~[\example{}:~\ref{TFF_Finite.s}]: \\
\hspace*{0.5cm}\smalltt{loves(d\_garfield) = d\_garfield}\\
The predicate mappings are literals, e.g.,~[\example{}:~\ref{TFF_Finite.s}]: \\
\hspace*{0.5cm}\smalltt{\tildenot{}\,owns(d\_jon,d\_nermal)}

The interpretation-formulae are preceded by the necessary type declarations:
\begin{packed_itemize}
\item The types in the problem.
\item The types of the domains.
\item The types of the domain elements.
\end{packed_itemize}

[\example{}:~\ref{TFF_Finite.s.p}] shows the verification problem for 
the model.

\subsection{Finite Interpretations with Domain Types}
\label{TarskianNewFiniteTypes}

Reusing the problem types for domain elements can get murky, e.g.,~when quantification over only 
the domain elements is intended. 
It is clearer to use new types for domain elements, and it becomes necessary when the domain 
elements are complex or of a defined type such as \smalltt{\$int} (see 
Section~\ref{TarskianInfinite}).
As noted in the introduction to Section~\ref{NewTarskian}, applying the problem's function and 
predicate symbols to domain elements that are defined with different types to the problem's types 
requires using type-promotion bijections to ``convert'' domain elements to terms of the 
corresponding problem type.

[\example{}:~\ref{TFF_Finite.p}] shows a TFF problem that has a 
countermodel with finite domains, which is shown in 
[\example{}:~\ref{TFF_Finite_SeparateDomains.s}].
[\example{}:~\ref{THF_Finite.p}] shows a THF problem that has a 
countermodel with finite domains, which is shown in
[\example{}:~\ref{THF_Finite.s}].

Each domain specification is a conjunction of:
\begin{packed_itemize}
\item Specification of the domain for each problem type, as a formula that makes the 
      type-promotion function a surjection, 
      e.g.,~[\example{}:~\ref{TFF_Finite_SeparateDomains.s}]:\\
      \hspace*{0.5cm}\smalltt{! [H: human] : ? [DH: d\_human] : H = d2human(DH)}\\
      \hspace*{0.5cm}\smalltt{! [C: cat] : ? [DC: d\_cat] : C = d2cat(DC)}\\
      and [\example{}:~\ref{THF_Finite.s}]:\\
      \hspace*{0.5cm}\smalltt{! [B: beverage] : ? [DB: d\_beverage] : ( B = ( d2beverage @ DB ) )}\\
      \hspace*{0.5cm}\smalltt{! [S: syrup] : ? [DS: d\_syrup] : ( S = ( d2syrup @ DS ) )}
\item Specification of the domain elements as a universally quantified disjunction of equalities 
      whose right-hand sides are the domain elements, 
      e.g.,~[\example{}:~\ref{TFF_Finite_SeparateDomains.s}]:\\
      \hspace*{0.5cm}\smalltt{! [DH: d\_human] : ( DH = d\_jon )}\\
      \hspace*{0.5cm}\smalltt{! [DC: d\_cat]: ( DC = d\_garfield | DC = d\_arlene | DC = d\_nermal )}\\
      and [\example{}:~\ref{THF_Finite.s}]:\\
      \hspace*{0.5cm}\smalltt{! [DB: beverage] : ( DB = d\_coffee )}\\
      \hspace*{0.5cm}\smalltt{! [DS: d\_syrup] : ( DS = d\_vanilla )}
\item Specification of the distinctness of the domain elements, 
      e.g.,~[\example{}:~\ref{TFF_Finite_SeparateDomains.s}]:\\
      \hspace*{0.5cm}\smalltt{\$distinct(d\_garfield,d\_arlene,d\_nermal)}\\
      If each domain has only one element this is unnecessary [\example{}:~\ref{THF_Finite.s}].
\item A formula making the type-promotion function an injection, which together with the 
      surjectivity makes it a bijection, e.g.,~[\example{}:~\ref{TFF_Finite_SeparateDomains.s}]:\\
      \hspace*{0.5cm}\smalltt{! [DH1: d\_human,DH2: d\_human] : ( d2human(DH1) = d2human(DH2) => DH1 = DH2 )}\\
      \hspace*{0.5cm}\smalltt{! [DC1: d\_cat,DC2: d\_cat] : ( d2cat(DC1) = d2cat(DC2) => DC1 = DC2 )}\\
      and [\example{}:~\ref{THF_Finite.s}]:\\
      \hspace*{0.5cm}\smalltt{! [DB1: d\_beverage,DB2: d\_beverage] :}\\
      \hspace*{0.8cm}\smalltt{( ( ( d2beverage @ DB1 ) = ( d2beverage @ DB2 ) ) => ( DB1 = DB2 ) )}\\
      \hspace*{0.5cm}\smalltt{! [DS1: d\_syrup,DS2: d\_syrup] :}\\
      \hspace*{0.8cm}\smalltt{( ( ( d2syrup @ DS1 ) = ( d2syrup @ DS2 ) ) => ( DS1 = DS2 ) )}
\end{packed_itemize}

The function mappings are equalities between type-promoted domain elements and functions applied 
to type-promoted domain elements, e.g.,~[\example{}:~\ref{TFF_Finite_SeparateDomains.s}]:\\
\hspace*{0.5cm}\smalltt{jon = d2human(d\_jon)}\\
\hspace*{0.5cm}\smalltt{loves(d2cat(d\_garfield)) = d2cat(d\_garfield)}\\
and [\example{}:~\ref{THF_Finite.s}]:\\
\hspace*{0.5cm}\smalltt{coffee = ( d2beverage @ d\_coffee )}\\
\hspace*{0.5cm}\smalltt{( heat @ ( d2beverage @ d\_coffee ) ) = ( d2beverage @ d\_coffee )}\\
The predicate mappings are literals, 
e.g.,~[\example{}:~\ref{TFF_Finite_SeparateDomains.s}]:\\
\hspace*{0.5cm}\smalltt{\tildenot{}\,owns(d2human(d\_jon),d2cat(d\_nermal))}\\
and [\example{}:~\ref{THF_Finite.s}]:\\
\hspace*{0.5cm}\smalltt{hot @ ( d2beverage @ d\_coffee )}\\
The mappings of higher-order function symbols are lambda abstractions, 
e.g.,~[\example{}:~\ref{THF_Finite.s}]:\\
\hspace*{0.5cm}\smalltt{mix = ( \textasciicircum~[F: syrup > beverage] : ( d2beverage @ d\_coffee ) )}\\
\hspace*{0.5cm}\smalltt{heated\_mix = ( \textasciicircum~[F: syrup > beverage] : ( d2beverage @ d\_coffee ) )}\\
The lambda abstractions are convenient because the interpretation of symbols that take a function 
argument must specify the interpretation for every such function. 
This differs from ground domain elements that can simply be assigned a name. 
A function is entirely characterised by its values on all inputs, so specifying a function-type 
element necessarily involves defining its behaviour, which is expressed as a lambda term. 
For infinite base types such as \smalltt{\$int}, the function space cannot be enumerated, 
providing an additional reason for using lambda abstractions, but the use of lambda abstractions 
is required regardless of cardinality.

The TFF and THF interpretation-formulae are preceded by the necessary type declarations:
\begin{packed_itemize}
\item The types in the problem.
\item The types of the domains.
\item The types of the domain elements.
\item The types of the type-promotion functions.
\end{packed_itemize}

[\example{}:~\ref{TFF_Finite_SeparateDomains.s.p}, 
\ref{THF_Finite.s.p}] show the verification problems for the models.
Sometimes higher-order formulae have models that can be expressed in TFF, which is preferable for
presentation and readability. 
However, since the model and its verification problem must be in the same language, and 
verification obligations might require higher-order reasoning, both must be stated in THF.

\subsection{Infinite Interpretations}
\label{TarskianInfinite}

Interpretations can have infinite domains formed from terms or defined types such as 
\smalltt{\$int}.
[\example{}:~\ref{TFF_Infinite.p}] shows a TFF problem that has a 
model with an infinite domain.
[\example{}:~\ref{TFF_Peano.s}] shows the model with an infinite term 
domain, and [\example{}:~\ref{TFF_Integer.s}] shows the model with an
\smalltt{\$int} domain.

Each domain specification is a conjunction of:
\begin{packed_itemize}
\item The domain for each problem type, as a formula that makes the type-promotion function a 
      surjection, e.g.,~[\example{}:~\ref{TFF_Peano.s}]:\\
      \hspace*{0.5cm}\smalltt{! [P: person] : ? [I: peano] : ( P = peano2person(I) )}\\
      and [\example{}:~\ref{TFF_Integer.s}]:\\
      \hspace*{0.5cm}\smalltt{! [P: person] : ? [I: \$int] : P = int2person(I)}\\
      If the problem type and the domain type are the same defined type, e.g.,~both are 
      \smalltt{\$int}, this is unnecessary.
\item Specification of the domain elements as an existentially quantified formula that captures an
      infinite disjunction of equalities, e.g.,~[\example{}:~\ref{TFF_Peano.s}]:\\
      \hspace*{0.5cm}\smalltt{! [I: peano] : ( I = zero | ? [P: peano] : I = s(P) )} \\
      If the domain is a defined type such as \smalltt{\$int} this is unnecessary
      [\example{}:~\ref{TFF_Integer.s}].
\item Specification of the distinctness of the domain elements, unless implicit from their type.
      e.g.,~[\example{}:~\ref{TFF_Peano.s}]:\\
      \hspace*{0.5cm}\smalltt{( peano\_less(I1,I2) => I1 != I2 )}\\
      If the domain is a defined type such as \smalltt{\$int} this is unnecessary
      [\example{}:~\ref{TFF_Integer.s}].
\item A formula making the type-promotion function an injection, which together with the 
      surjectivity makes it a bijection, e.g.,~[\example{}:~\ref{TFF_Peano.s}]:\\
      \hspace*{0.5cm}\smalltt{! [I1: peano,I2: peano] : ( peano2person(I1) = peano2person(I2) => I1 = I2 )}\\
      and [\example{}:~\ref{TFF_Integer.s}]:\\
      \hspace*{0.5cm}\smalltt{! [I1: \$int,I2: \$int] : ( int2person(I1) = int2person(I2) => I1 = I2 )}
\end{packed_itemize}

The function mappings are universally quantified equalities between type-promoted domain 
elements and functions applied to type-promoted domain elements, 
e.g.,~[\example{}:~\ref{TFF_Peano.s}]:\\
\hspace*{0.5cm}\smalltt{! [I: peano] : child\_of(peano2person(I)) = peano2person(s(I))}\\
and [\example{}:~\ref{TFF_Integer.s}]:\\
\hspace*{0.5cm}\smalltt{! [I: \$int] : child\_of(int2person(I)) = int2person(\$sum(I,1))}\\
The predicate mappings are universally quantified formulae, 
e.g.,~[\example{}:~\ref{TFF_Peano.s}]:\\
\hspace*{0.5cm}\smalltt{! [A: peano,D: peano] :}\\
\hspace*{0.9cm}\smalltt{( is\_descendant(peano2person(A),peano2person(D)) <=> peano\_less(A,D) )}\\
and [\example{}:~\ref{TFF_Integer.s}]:\\
\hspace*{0.5cm}\smalltt{! [A: \$int,D: \$int] :}\\
\hspace*{0.9cm}\smalltt{( is\_descendant(int2person(A),int2person(D))}\\
\hspace*{0.5cm}\smalltt{<=> peano\_less(A,D) )}\\
The interpretation of function and predicate symbols is not given for argument tuples
with negative integers.

The interpretation-formulae are preceded by the necessary type declarations:
\begin{packed_itemize}
\item The types in the problem.
\item The types of the domains.
\item The types of the domain elements.
\item The types of the type-promotion functions.
\end{packed_itemize}

[\example{}:~\ref{TFF_Peano.s.p}, \ref{TFF_Integer.s.p}] show the 
verification problems for the models.

\subsection{Interpretations with Multiple Interpretation-formulae}
\label{NewTarskianSplit}

The interpretation-formulae described thus far have all the information about an interpretation 
in a single interpretation-formula.
In some situations it is useful to split the various components into separate 
interpretation-formulae, e.g.,~separating the domains from the symbol mappings.
The TPTP format offers splitting in a flexible way, at medium and fine grained levels, using 
annotated formula subroles.

At the medium grained level, the \smalltt{interpretation} role can be extended with the subroles
\smalltt{domains} and \smalltt{mappings}.
Two (or more) interpretation-formulae are used, each containing the corresponding parts of the 
interpretation.
[\example{}:~\ref{FOF_Finite_Medium.s}, \ref{TFF_Finite_Medium.s}]
show examples of splitting the interpretation-formulae in the FOF and TFF 
interpretations [\example{}:~\ref{FOF_Finite.s}, \ref{TFF_Finite.s}] 
respectively, which separate the domain specifications from the symbol mappings.
For example, [\example{}:~\ref{TFF_Finite_Medium.s}]:\\
\hspace*{0.5cm}\smalltt{tff(garfield\_domains,interpretation-domains,}\emph{the domains}\smalltt{).} \\
contains the information about the two domains, and the interpretation-formula: \\
\hspace*{0.5cm}\smalltt{tff(garfield\_mappings,interpretation-mappings,}\emph{the mappings}\smalltt{).} \\
contains the symbol mappings.
Note that each role and role-subrole pair can be used multiple times according to need, 
e.g.,~[\example{}:~\ref{FOF_Finite_Medium.s}] there can be separate
\smalltt{interpretation-mappings} interpretation-formulae, one for functions and one for 
predicates.

At the fine grained level, interpretation-formulae can split the individual domains and symbol 
mappings, with the \smalltt{domains} and \smalltt{mappings} subroles given arguments to indicate
which domain or symbol mapping is recorded.
For \smalltt{domains} the arguments of the subrole are the domain in the problem and the domain 
in the interpretation.
For \smalltt{mappings} the arguments are the symbol and its result domain type.
[\example{}:~\ref{FOF_Finite_Fine.s}, \ref{TFF_Finite_Fine.s}] show 
examples of splitting the interpretation-formulae in
[\example{}:~\ref{FOF_Finite.s}, \ref{TFF_Finite.s}] respectively.
For example, [\example{}:~\ref{TFF_Finite_Fine.s}]:\\
\hspace*{0.5cm}\smalltt{tff(garfield\_domain\_human,interpretation-domains(human,d\_human)} \\
contains the \smalltt{d\_human} domain specification, and: \\
\hspace*{0.5cm}\smalltt{tff(garfield\_domain\_cat,interpretation-domains(cat,d\_cat)} \\
contains the \smalltt{d\_cat} domain specification.

The medium and fine grained splitting can be mixed.
[\example{}:~\ref{THF_Finite_Medium.s}] shows an example of mixed 
splitting of [\example{}:~\ref{THF_Finite.s}].
For example, the domains are specified separately~[\example{}:~\ref{THF_Finite_Medium.s}]: \\
\hspace*{0.5cm}\smalltt{thf(hot\_coffee\_beverage,interpretation-domains(beverage,d\_beverage),}\\
\hspace*{0.9cm}\emph{beverage domain}\smalltt{).} \\
\hspace*{0.5cm}\smalltt{thf(hot\_coffee\_syrup,interpretation-domains(syrup,d\_syrup),}\emph{syrup domain}\smalltt{).} \\
while all the symbol mappings are in: \\
\hspace*{0.5cm}\smalltt{thf(hot\_coffee,interpretation-mappings,}\emph{the mappings}\smalltt{).}

\subsection{Interpretations Compacted by Quantification}
\label{NewTarskianCompact}

The full logical language can be used in interpretation-formulae to provide compaction.
[\example{}:~\ref{TFF_Finite_Compact.s}] shows an example compacting 
the interpretation-formula \smalltt{garfield\_mapping\_loves} in
[\example{}:~\ref{TFF_Finite.s}].
For example, universal quantification can be used to map \smalltt{loves} to \smalltt{d\_garfield} 
for all \smalltt{d\_cats}~[\example{}:~\ref{TFF_Finite_Compact.s}]: \\
\hspace*{0.5cm}\smalltt{!\,[DC:\,d\_cat]\,:\,(\,loves(d2cat(DC))\,=\,d2cat(d\_garfield)\,)}

\subsection{HI-formulae and Saturations}
\label{NewHerbrand}

HI-formulae determine sets of Herbrand interpretations.
Interpretation-formulae can be used for these, although the actual Herbrand interpretations are
not easy to extract.
To indicate that the interpretation-formulae are intended to determine Herbrand interpretations 
they are given the subrole \smalltt{herbrand}.
[\example{}:~\ref{FOF_Finite.p}] shows a problem that has a HI-formulae
model formed from predicate definitions over the term algebra, which is shown in 
[\example{}:~\ref{FOF_Formulae.s}].
[\example{}:~\ref{FOF_Finite.p}] also has a saturation, which is shown 
in [\example{}:~\ref{FOF_Saturation.s}].
[\example{}:~\ref{FOF_Formulae.s.p}, \ref{FOF_Saturation.s.p}] show 
the corresponding verification problems.

\subsection{Visualization of Interpretations}
\label{IIV}

\begin{figure}[htbp]
\centering
\includegraphics[width=0.75\textwidth]{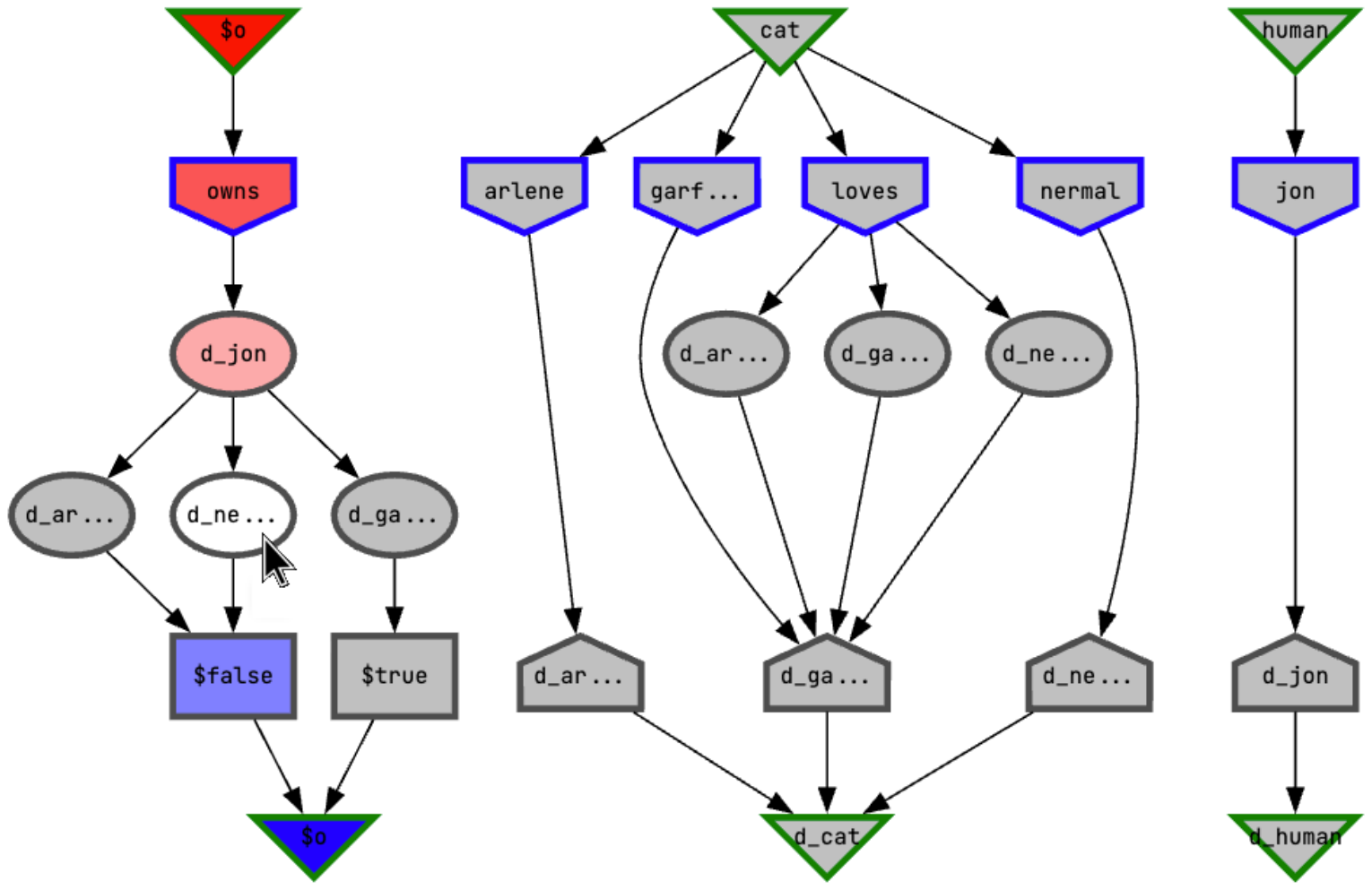}
\caption{IIV view of [\example{}:~\ref{TFF_Finite_SeparateDomains.s}]}
\label{IIVPicture}
\end{figure}

Finite Tarskian interpretations with domain types, as described in 
Section~\ref{TarskianNewFiniteTypes}, can be visualised in the Interactive Interpretation Viewer 
(IIV)~\cite{MS23-Poster}.
Figure~\ref{IIVPicture} shows the view of the model 
in [\example{}:~\ref{TFF_Finite_SeparateDomains.s}].
The cursor is hovering over the \smalltt{d\_nermal} node in the interpretation of the
\smalltt{owns} predicate applied to \smalltt{d\_jon} and \smalltt{d\_nermal}, which is 
interpreted as \smalltt{\$false}.
The \smalltt{\$o} triangle above the \smalltt{owns} inverted house shows that \smalltt{owns} has
a boolean result in the problem, and the \smalltt{\$o} triangle below \smalltt{\$false} box shows 
that the interpretation value is boolean.
IIV is available in the SystemOnTSTP web interface.

\section{The Format for Kripke Interpretations}
\label{NewKripke}
 
The TPTP format for Kripke interpretations also uses interpretation-formulae.
As was noted in Section~\ref{Have}, Kripke interpretations can have either a finite or an
infinite number of worlds, and the interpretations in a world can be Tarskian or 
Herbrand.
The TPTP format for Kripke interpretation-formulae adds four defined symbols:
\begin{packed_itemize}
\item A defined type \smalltt{\$world} for the worlds of the interpretation.
      Different constants of type \smalltt{\$world} are known to be unequal.
\item A defined predicate \smalltt{\$accessible\_world} of type {\tt (\$world\,*\,\$world)\,>\,\$o}
      to specify the accessibility relation between worlds.
\item A defined constant \smalltt{\$local\_world} of type \smalltt{\$world} to specify 
      the world in which a local (the default) conjecture is to be proved. 
\item A defined predicate \smalltt{\$in\_world} of type \smalltt{(\$world\,*\,\$o)\,>\,\$o} to 
      specify the interpretations in the worlds.
\end{packed_itemize}

A single Kripke interpretation-formula is a conjunction of:
\begin{packed_itemize}
\item A specification of the worlds.
\item Explication of the distinctness of the worlds (by definition different worlds are known to
      be distinct, but no ATP systems know that yet, so it's necessary to encode the
      distinctness explicitly using inequalities or the \smalltt{\$distinct} predicate).
\item The accessibility relation.
\item Specification of the local world if any.
\item For each world, its Tarskian interpretation (also in the TPTP format for interpretations),
      augmented by existential subformulae that require the existence of the world's domain 
      elements -- these are necessary because the semantics is that of modal logic (see
      Sections~\ref{NewKripkeFiniteFinite} and~\ref{NewKripkeSplitCompact} for examples).
\end{packed_itemize}
The interpretation-formulae are preceded by the necessary type declarations:
\begin{packed_itemize}
\item The types for the Tarskian interpretations in the worlds.
\item The worlds declared of type \smalltt{\$world}, e.g.,~\smalltt{w1: \$world}.
\end{packed_itemize}

The \smalltt{logic} specification of the problem is included to specify that the interpretation is
for formulae in that logic.
This information is needed when processing an interpretation, e.g.,~in verification (see 
Section~\ref{Verification}). 
The interpretation-formula does not provide this information because it underspecifies the 
logic in use, e.g.,~it's usually not possible to see whether the interpretation exemplifies modal 
system K or modal system S5 -- in both cases the interpretation could interpret the accessibility 
relation as an equivalence relation (this is required for S5 but it is also OK for K).

[\example{}:~\ref{NXF_Finite-Finite-Global.s}, \ref{NXF_Finite-Finite-Local.s}] illustrate the 
components of finite Kripke interpretations in a single interpretation-formula, as explained in 
Sections~\ref{NewKripkeFiniteFinite}.
[\example{}:~\ref{NXF_Finite-Finite-Global_Medium.s},
\ref{NXF_Finite-Finite-Global_Fine.s}] illustrate Kripke interpretations split 
over multiple interpretation-formulae, as explained in Section~\ref{NewKripkeSplitCompact}.
[\example{}:~\ref{NXF_Finite-Finite-Global_Compact.s}] illustrate
a Kripke interpretations that is compacted using quantification, as explained in 
Section~\ref{NewKripkeSplitCompact}.
These examples illustrate the possibilities for writing Kripke interpretations in the TPTP format,
but are not intended to be exhaustive.
In particular, Kripke interpretations with an infinite number of worlds are not considered.

\subsection{Finite-Finite Kripke Interpretations}
\label{NewKripkeFiniteFinite}

[\example{}:~\ref{NXF_Finite-Finite-Global.p}] shows a NXF problem 
that has a countermodel with finite worlds, each of which contains a finite Tarskian 
interpretation, which is shown in 
[\example{}:~\ref{NXF_Finite-Finite-Global.s}].
The problem has global axioms and a local conjecture.
The countermodel has three worlds [\example{}:~\ref{NXF_Finite-Finite-Global.s}]: \\
\hspace*{0.5cm}\smalltt{!\,[W:\,\$world]\,:\,(\,W\,=\,w1\,|\,W\,=\,w2\,|\,W\,=\,w3\,)} \\
\hspace*{0.5cm}\smalltt{\$distinct(w1,w2,w3)} \\
with the accessibility relation: \\
\hspace*{0.9cm}\smalltt{\$accessible\_world(w1,w1) \& \$accessible\_world(w2,w2)}\\
\hspace*{0.5cm}\smalltt{\& \$accessible\_world(w1,w2) \& \$accessible\_world(w2,w3)}\\
\hspace*{0.5cm}\smalltt{\& \$accessible\_world(w3,w1)}\\
\hspace*{0.5cm}\smalltt{\& \tildenot{}\,\$accessible\_world(w1,w3) \& \tildenot{}\,\$accessible\_world(w2,w1)}\\
\hspace*{0.5cm}\smalltt{\& \tildenot{}\,\$accessible\_world(w3,w2) \& \tildenot{}\,\$accessible\_world(w3,w3)}\\
The conjecture is disproved in a local world [\example{}:~\ref{NXF_Finite-Finite-Global.s}]: \\
\hspace*{0.5cm}\smalltt{\$local\_world = w1} \\
The finite Tarskian interpretations for the worlds are provided in the format described in 
Section~\ref{NewTarskian}: \\
\hspace*{0.5cm}\smalltt{\$in\_world(w1,}\emph{the Tarskian interpretation}\smalltt{)} \\
\hspace*{0.5cm}\smalltt{\$in\_world(w2,}\emph{the Tarskian interpretation}\smalltt{)} \\
\hspace*{0.5cm}\smalltt{\$in\_world(w3,}\emph{the Tarskian interpretation}\smalltt{)} \\
Note the augmentation of the Tarskian interpretations with subformulae such as:\\
\hspace*{0.5cm}\smalltt{?\,[CD:\,child\_d]\,:\,CD\,=\,child\_1}\\
The quantification semantics is not classical, but instead that of modal logic, i.e., the 
quantification is over the domain elements that exist in the world.
That subformula requires that the domain element \smalltt{child\_1} exists in the world.
[\example{}:~\ref{NXF_Finite-Finite-Global.s.p}] shows the verification problem for the model.

[\example{}:~\ref{NXF_Finite-Finite-Local.p}] shows a NXF problem 
that has both local and global axioms.
It has a countermodel with finite worlds each of which contains a finite Tarskian interpretation, 
which is shown in [\example{}:~\ref{NXF_Finite-Finite-Local.s}].
This example illustrates how a local axiom (with role \smalltt{axiom-local}) is satisfied in 
only the local world (\smalltt{w1}).
[\example{}:~\ref{NXF_Finite-Finite-Local.s.p}] shows the verification 
problem for the model.

\subsection{Kripke Interpretations Split and Compacted}
\label{NewKripkeSplitCompact}

Kripke interpretation-formulae can be split to separate the various components of the 
interpretation.
The \smalltt{interpretation} role can be extended with the subroles \smalltt{worlds},
\smalltt{domains}, and \smalltt{mappings}.
[\example{}:~\ref{NXF_Finite-Finite-Global_Medium.s}] shows a medium 
grained split of [\example{}:~\ref{NXF_Finite-Finite-Global.s}].
An interpretation-formula with the role \smalltt{interpretation-worlds} contains information about 
the worlds of the interpretation.
For example, one interpretation-formula contains the information about the three 
worlds~[\example{}:~\ref{NXF_Finite-Finite-Global_Medium.s}]: \\
\hspace*{0.5cm}\smalltt{tff(people\_worlds,interpretation-worlds,}\emph{world information}\smalltt{).}

Splitting can be done in a flexible way, using the fine grained splitting available for
Tarskian interpretation-formulae.
For example, the interpretation-formula~[\example{}:~\ref{NXF_Finite-Finite-Global_Fine.s}]: \\
\hspace*{0.5cm}\smalltt{tff(child\_domains,interpretation-domains(child,child\_d),}\emph{child domain}\smalltt{).} \\
contains the information about the domain \smalltt{child\_d}, the interpretation-formula: \\
\hspace*{0.5cm}\smalltt{tff(people\_mappings,interpretation-mappings(charly,child\_d)}\emph{charly mapping}\smalltt{).} \\
contains the mapping for the \smalltt{charly} constant, and the interpretation-formula: \\
\hspace*{0.5cm}\smalltt{tff(rains\_mappings,interpretation-mappings(rains,\$o),}\emph{rains mapping}\smalltt{).} \\
contains the mapping for the \smalltt{rains} predicate.

In the same way that Tarskian interpretation-formulae can be compacted, Kripke
interpretation-formulae can be compacted, e.g.,~by using universal quantification over the
worlds to factor out common elements of the Tarskian interpretations in the worlds.
For example, the interpretation-formula [\example{}:~\ref{NXF_Finite-Finite-Global_Compact.s}]: \\
\hspace*{0.5cm}\smalltt{tff(people\_domains,interpretation-domains,} \\
\hspace*{0.9cm}\smalltt{! [W: \$world] :} \\
\hspace*{1.3cm}\smalltt{\$in\_world(W,}\emph{the domain specification}\smalltt{).} \\
contains the information about the domains that are the same in all worlds 
[\example{}:~\ref{NXF_Finite-Finite-Global.s}], and the interpretation-formula
[\example{}:~\ref{NXF_Finite-Finite-Global_Compact.s}]: \\
\hspace*{0.5cm}\smalltt{tff(charly\_mappings,interpretation-mappings(charly,child\_d),}\\
\hspace*{0.9cm}\smalltt{! [W: \$world] :} \\
\hspace*{1.3cm}\smalltt{\$in\_world(W,} \\
\hspace*{1.7cm}\smalltt{( charly = d2child(child\_1) )) ).} \\
contains the mapping for the \smalltt{charly} constant that is the same in all worlds
[\example{}:~\ref{NXF_Finite-Finite-Global.s}], and the interpretation-formula
[\example{}:~\ref{NXF_Finite-Finite-Global_Compact.s}]: \\
\hspace*{0.5cm}\smalltt{tff(rains\_mappings,interpretation-mappings(rains,\$o),} \\
\hspace*{0.5cm}\smalltt{! [W: \$world] :} \\
\hspace*{0.9cm}\smalltt{\$in\_world(W,} \\
\hspace*{1.3cm}\smalltt{rains) ).} \\
contains the predicate mapping for \smalltt{rains} that is the same in all worlds
[\example{}:~\ref{NXF_Finite-Finite-Global.s}].

\subsection{Kripke Model Verification}
\label{KripkeVerification}

Kripke models written as interpretation-formulae can be verified using the theorem proving 
approach described in Section~\ref{Verification}.\footnote{%
To date we have done this for NXF problem formulae and TXF interpretation-formulae. 
Extension to NHF should be possible.}
An NXF verification problem is built from the NXF problem formulae and the TXF 
interpretation-formulae.
Although the verification problem includes the non-classical connectives in the problem
formulae, it is not necessary to provide a full logic specification (as in 
[\example{}:~\ref{NXF_Finite-Finite-Global.p}, 
\ref{NXF_Finite-Finite-Local.p}]) because the logical setting is captured in 
the interpretation-formulae.
The verification problem is thus not in the same logic as the problem formulae, but rather
some kind of hybrid logic~\cite{Bra11}, which is indicated by the special logic specification
value \smalltt{\$\$fomlModel}.
[\example{}:~\ref{NXF_Finite-Finite-Global.s.p}] shows the NXF 
verification problem for the NXF problem
[\example{}:~\ref{NXF_Finite-Finite-Global.p}] and its countermodel 
in [\example{}:~\ref{NXF_Finite-Finite-Global.s}].
[\example{}:~\ref{NXF_Finite-Finite-Local.s.p}] shows the NXF 
verification problem for the problem in 
[\example{}:~\ref{NXF_Finite-Finite-Local.p}] and its countermodel in 
[\example{}:~\ref{NXF_Finite-Finite-Local.s}].

In the NXF verification problem the Kripke interpretation-formulae are given the \smalltt{axiom} 
role, and each of the problem axioms and the negated problem conjecture are given the 
\smalltt{conjecture} role with a subrole of \smalltt{local} or \smalltt{global} according to 
their role in the problem.
For example, the axiom~[\example{}:~\ref{NXF_Finite-Finite-Global.p}]: \\
\hspace*{0.5cm}\smalltt{tff(a1,axiom} \\
\hspace*{0.9cm}\smalltt{! [C: child] : \tildenot{}\,( \tildenot{}\,quiet(C) \& ? [A: adult] : sleepy(A) ) ).}\\
is global (the default), so the NXF verification conjecture has the \smalltt{global} subrole
[\example{}:~\ref{NXF_Finite-Finite-Global.s.p}]: \\
\hspace*{0.5cm}\smalltt{tff(a1,conjecture-global,} \\
The conjecture of the problem is local (the default) 
[\example{}:~\ref{NXF_Finite-Finite-Global.p}]: \\
\hspace*{0.5cm}\smalltt{tff(c,conjecture,} \\
\hspace*{0.9cm}\smalltt{\{\$possible\}\,@\,(\,\tildenot{}\,rains\,)\,).} \\
so the NXF verification (negated) conjecture has the \smalltt{local} subrole
[\example{}:~\ref{NXF_Finite-Finite-Global.s.p}]: \\
\hspace*{0.5cm}\smalltt{tff(c,conjecture-local,}\emph{verification obligation}\smalltt{).}

The NXF verification problem is embedded into TH0 using the ATFLET tool.
ATFLET recognizes the special logic specification and the type and predicates, and combines
the NXF conjectures into a single TH0 conjecture.
The resultant TH0 problem is discharged using a trusted TH0 ATP system.

\subsection{Visualization of Interpretations}
\label{IKV}

\begin{figure}[h]
\centering
\includegraphics[width=0.75\textwidth]{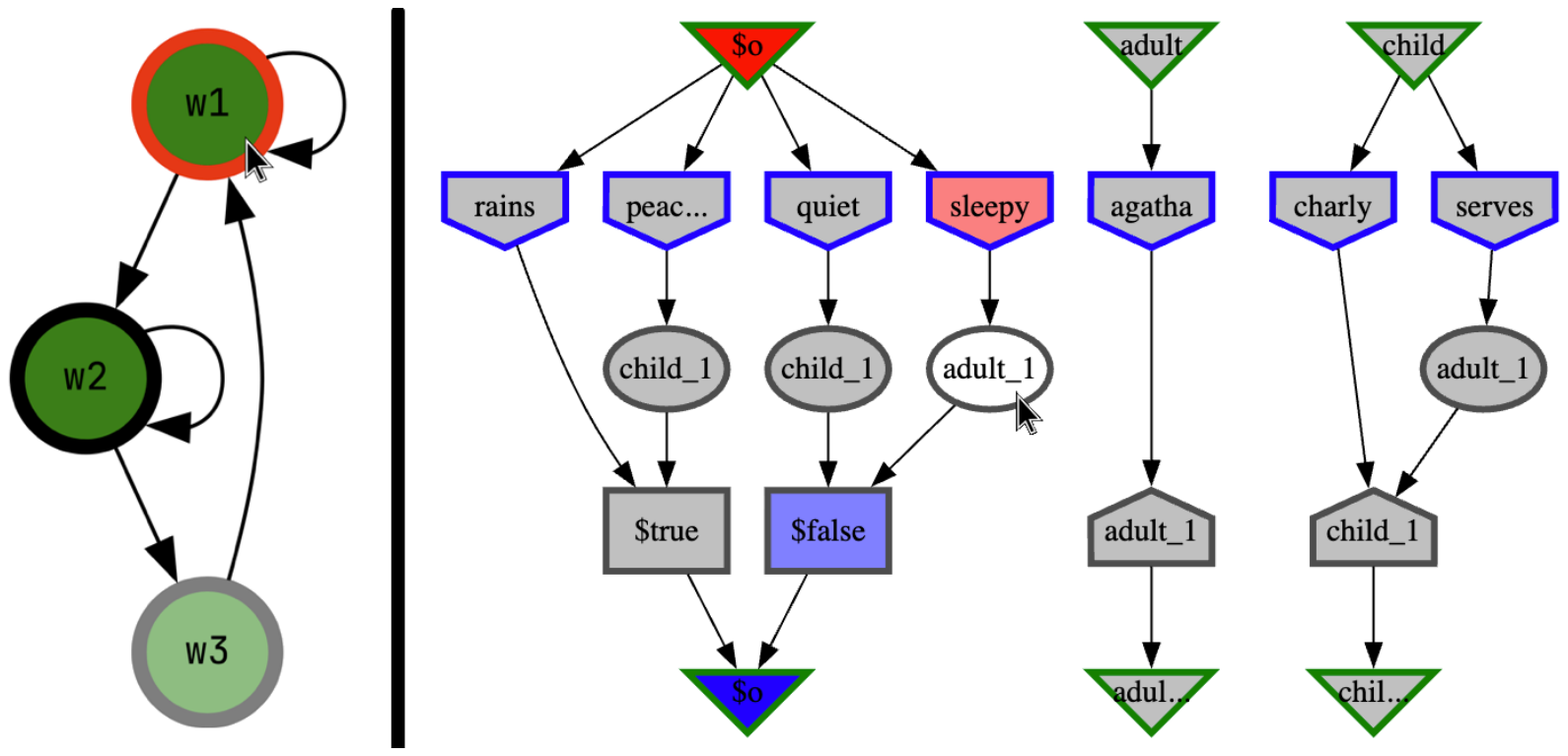}
\caption{IIV view of the Kripke model in [\example{}:~\ref{NXF_Finite-Finite-Global.s}]}
\label{IKVPicture}
\end{figure}

Finite Kripke interpretations with domain types, as described in 
Section~\ref{NewKripkeFiniteFinite}, can be visualised in the Interactive Interpretation Viewer 
(IIV).
The left hand side of Figure~\ref{IKVPicture} shows the worlds of the model in 
[\example{}:~\ref{NXF_Finite-Finite-Global.s}].
The cursor is hovering over the \smalltt{w1} world, and the worlds are coloured in lighter
shades according to the number of accessibility relations that have to be traversed from
the selected world.
Clicking on a world displays,the Tarskian interpretation in the world, shown in the
right hand side of Figure~\ref{IKVPicture}.
IIV is available in the SystemOnTSTP web interface.

\section{Conclusion}
\label{Conclusion}

This paper describes the TPTP format for representing interpretations.
It provides a background survey that helped ensure that the representation format is adequate
for different types of interpretations, including Tarskian, Herbrand, and Kripke interpretations.
Some tools that support processing the format have been noted.

The TPTP format provides numerous options and features, and it is expected that only the 
basic features will be adopted initially. 
For ATP systems that already output the old TPTP format for finite Tarskian interpretations,
all that is needed is to replace the {\tt fi\_domain}, {\tt fi\_functors}, and {\tt fi\_predicates}
roles by {\tt interpretation}. 
To be more informative, {\tt fi\_domain} can be replaced by {\tt interpretation-domains}, and 
{\tt fi\_functors} and {\tt fi\_predicates} can be replaced by {\tt interpretation\-mappings}.

Section~\ref{Need} lists four needs for interpretations: 
\begin{packed_enumerate}
\item Evaluability - the ability to evaluate a formula wrt an interpretation.
\item Tractability - the ability to evaluate a formula using some limited/reasonable amount of
      resources.
\item Verifiability - the ability to verify a model by evaluating all the formulae it claims
      to model as \textit{true}.
\item Comprehensibility - be sufficiently comprehensible to humans and machines to be useful in 
      applications.
\end{packed_enumerate}

In the light of the above (and hey, maybe there are more techniques than just those) we (well,
at least the first author) claim:
\begin{packed_itemize}
\item Finite interpretations represented by interpretation-formulae are typically evaluable,
      tractable, verifiable, and comprehensible.
\item Infinite interpretations represented by interpretation-formulae are evaluable, often 
      intractable, verifiable, and might not be comprehensible.
\item HI-formulae (in interpretation-formulae) are evaluable, can be tractable, verifiable, 
      and can be comprehensible.
\item Saturations (in interpretation-formulae) are evaluable, often intractable, verifiable, 
      and almost always incomprehensible.
\item The verifiability, tractability, and comprehensibility of Kripke interpretations vary
      between individual cases.
\item Use of the same language as the problem formulae for writing interpretation-formulae
      contributes to comprehensibility (but does not ensure it).
\end{packed_itemize}

This work will be extended to deal with more non-classical languages.
But for now we are pausing to allow ATP system developers to implement the format in their 
ATP systems.

\bibliographystyle{plain}
\bibliography{Bibliography.bib}
\appendix

\newpage
\section{FOF}
\label{FOF}

\subsection{FOF Problem with a Finite Countermodel
(\href{https://raw.githubusercontent.com/GeoffsPapers/InterpretationFormat/master/Examples/FOF_Finite.p}{Online})}
\label{FOF_Finite.p}
\begin{small}
\begin{verbatim}
%------------------------------------------------------------------------------
%----All (hu)men are created equal. John is a human. John got an F grade.
%----There is someone (a human) who got an A grade. An A grade is not 
%----equal to an F grade. Grades are not human. Therefore, it is not the 
%----case being created equal is the same as really being equal.

fof(all_created_equal,axiom,
    ! [H1,H2] :
      ( ( human(H1)
        & human(H2) )
     => created_equal(H1,H2) ) ).

fof(john,axiom,
    human(john) ).

fof(john_failed,axiom,
    grade_of(john) = f ).

fof(someone_got_an_a,axiom,
    ? [H] :
      ( human(H)
      & grade_of(H) = a ) ).

fof(distinct_grades,axiom,
    a != f ).

fof(grades_not_human,axiom,
    ! [G] : ~ human(grade_of(G)) ).

fof(equality_lost,conjecture,
    ! [H1,H2] :
      ( ( human(H1)
        & human(H2)
        & created_equal(H1,H2) )
    <=> H1 = H2 ) ).
%------------------------------------------------------------------------------
\end{verbatim}
\end{small}

\newpage
\subsection{FOF Finite Model for \ref{FOF_Finite.p}, Coarse Grained
(\href{https://raw.githubusercontent.com/GeoffsPapers/InterpretationFormat/master/Examples/FOF_Finite.s}{Online})}
\label{FOF_Finite.s}
\begin{small}
\begin{verbatim}
%------------------------------------------------------------------------------
fof(equality_lost,interpretation,
    ( ! [X] : ( X = "a" | X = "f" | X = "john" | X = "gotA") 
    & ( a = "a"
      & f = "f"
      & john = "john"
      & grade_of("a") = "a"
      & grade_of("f") = "a"
      & grade_of("john") = "f"
      & grade_of("gotA") = "a" ) 
    & ( ~ human("a")
      & ~ human("f")
      & human("john")
      & human("gotA")
      & ~ created_equal("a","a")
      & ~ created_equal("a","f")
      & ~ created_equal("a","john")
      & ~ created_equal("a","gotA")
      & ~ created_equal("f","a")
      & ~ created_equal("f","f")
      & ~ created_equal("f","john")
      & ~ created_equal("f","gotA")
      & ~ created_equal("john","a")
      & ~ created_equal("john","f")
      & created_equal("john","john")
      & created_equal("john","gotA")
      & ~ created_equal("gotA","a")
      & ~ created_equal("gotA","f")
      & created_equal("gotA","john")
      & created_equal("gotA","gotA") ) ) ).
%------------------------------------------------------------------------------
\end{verbatim}
\end{small}

\newpage
\subsection{FOF Verification Problem for \ref{FOF_Finite.p} and \ref{FOF_Finite.s}
(\href{https://raw.githubusercontent.com/GeoffsPapers/InterpretationFormat/master/Examples/FOF_Finite.s.p}{Online})}
\label{FOF_Finite.s.p}
\begin{small}
\begin{verbatim}
%------------------------------------------------------------------------------
fof(equality_lost,axiom,
    ( ! [X] : ( X = "a" | X = "f" | X = "john" | X = "gotA" )
    & a = "a"
    & f = "f"
    & john = "john"
    & grade_of("a") = "a"
    & grade_of("f") = "a"
    & grade_of("john") = "f"
    & grade_of("gotA") = "a"
    & ~ human("a")
    & ~ human("f")
    & human("john")
    & human("gotA")
    & ~ created_equal("a","a")
    & ~ created_equal("a","f")
    & ~ created_equal("a","john")
    & ~ created_equal("a","gotA")
    & ~ created_equal("f","a")
    & ~ created_equal("f","f")
    & ~ created_equal("f","john")
    & ~ created_equal("f","gotA")
    & ~ created_equal("john","a")
    & ~ created_equal("john","f")
    & created_equal("john","john")
    & created_equal("john","gotA")
    & ~ created_equal("gotA","a")
    & ~ created_equal("gotA","f")
    & created_equal("gotA","john")
    & created_equal("gotA","gotA") ) ).

fof(prove_formulae,conjecture,
    ( ! [H1,H2] :
        ( ( human(H1)
          & human(H2) )
       => created_equal(H1,H2) )
    & human(john)
    & grade_of(john) = f
    & ? [H] :
        ( human(H)
        & grade_of(H) = a )
    & a != f
    & ! [G] : ~ human(grade_of(G))
    & ~ ! [H1,H2] :
          ( ( human(H1)
            & human(H2)
            & created_equal(H1,H2) )
        <=> H1 = H2 ) ) ).
%------------------------------------------------------------------------------
\end{verbatim}
\end{small}

\newpage
\subsection{FOF Finite Model for \ref{FOF_Finite.p}, Medium Grained
(\href{https://raw.githubusercontent.com/GeoffsPapers/InterpretationFormat/master/Examples/FOF_Finite_Medium.s}{Online})}
\label{FOF_Finite_Medium.s}
\begin{small}
\begin{verbatim}
%------------------------------------------------------------------------------
fof(equality_lost_domain,interpretation-domains,
    ! [X] : ( X = "a" | X = "f" | X = "john" | X = "gotA") ).

fof(equality_lost_term_mappings,interpretation-mappings,
    ( a = "a"
    & f = "f"
    & john = "john"
    & grade_of("a") = "a"
    & grade_of("f") = "a"
    & grade_of("john") = "f"
    & grade_of("gotA") = "a" ) ).

fof(equality_lost_predicate_mappings,interpretation-mappings,
    ( ~ human("a")
    & ~ human("f")
    & human("john")
    & human("gotA") ).
    & ~ created_equal("a","john")
    & ~ created_equal("a","gotA")
    & ~ created_equal("f","john")
    & ~ created_equal("f","gotA")
    & ~ created_equal("john","a")
    & ~ created_equal("john","f")
    & ~ created_equal("gotA","a")
    & ~ created_equal("gotA","f")
    & ~ created_equal("a","a")
    & ~ created_equal("a","f")
    & ~ created_equal("f","a")
    & ~ created_equal("f","f")
    & created_equal("john","john")
    & created_equal("john","gotA")
    & created_equal("gotA","john")
    & created_equal("gotA","gotA") ).
%------------------------------------------------------------------------------
\end{verbatim}
\end{small}

\newpage
\subsection{FOF Finite Model for \ref{FOF_Finite.p}, Fine Grained
(\href{https://raw.githubusercontent.com/GeoffsPapers/InterpretationFormat/master/Examples/FOF_Finite_Fine.s}{Online})}
\label{FOF_Finite_Fine.s}
\begin{small}
\begin{verbatim}
%------------------------------------------------------------------------------
fof(equality_lost_domain,interpretation-domains($i,$i),
    ! [X] : ( X = "a" | X = "f" | X = "john" | X = "gotA") ).

fof(equality_lost_a,interpretation-mappings(a,$i),
    a = "a" ).

fof(equality_lost_f,interpretation-mappings(f,$i),
    f = "f" ).

fof(equality_lost_john,interpretation-mappings(john,$i),
    john = "john" ).

fof(equality_lost_grade_of,interpretation-mappings(grade_of,$i),
    ( grade_of("a") = "a"
    & grade_of("f") = "a"
    & grade_of("john") = "f"
    & grade_of("gotA") = "a" ) ).

fof(equality_lost_human,interpretation-mappings(human,$o),
    ( ~ human("a")
    & ~ human("f")
    & human("john")
    & human("gotA") ).

fof(equality_lost_created_equal,interpretation-mappings(created_equal,$o),
    ( ~ created_equal("a","john")
    & ~ created_equal("a","gotA")
    & ~ created_equal("f","john")
    & ~ created_equal("f","gotA")
    & ~ created_equal("john","a")
    & ~ created_equal("john","f")
    & ~ created_equal("gotA","a")
    & ~ created_equal("gotA","f") ).
    & ~ created_equal("a","a")
    & ~ created_equal("a","f")
    & ~ created_equal("f","a")
    & ~ created_equal("f","f")
    & created_equal("john","john")
    & created_equal("john","gotA")
    & created_equal("gotA","john")
    & created_equal("gotA","gotA") ).
%------------------------------------------------------------------------------
\end{verbatim}
\end{small}

\newpage
\subsection{FOF Formula Model for \ref{FOF_Finite.p}
(\href{https://raw.githubusercontent.com/GeoffsPapers/InterpretationFormat/master/Examples/FOF_Formulae.s}{Online})}
\label{FOF_Formulae.s}
\begin{small}
\begin{verbatim}
%------------------------------------------------------------------------------
fof(created_equal,interpretation-herbrand,
    ! [X0,X1] : ( created_equal(X0,X1) <=> $true ) ).

fof(human,interpretation-herbrand,
    ! [X0] : ( human(X0) <=> ( X0 != "f" & X0 != "a" ) ) ).

fof(john,interpretation-herbrand,
    ! [X0] : ( X0 = john <=> X0 = "john" ) ).

fof(grade_of,interpretation-herbrand,
    ! [X0,X1] :
      ( X0 = grade_of(X1)
    <=> ( ( X0 = "f" & X1 != "gotA" )
        | ( X0 = "a" & X1 = "gotA" ) ) ) ).

fof(f,interpretation-herbrand,
    ! [X0] : ( X0 = f <=> X0 = "f" ) ).

fof(a,interpretation-herbrand,
    ! [X0] : ( X0 = a <=> X0 = "a" ) ).
%------------------------------------------------------------------------------
\end{verbatim}
\end{small}

\newpage
\subsection{FOF Saturation for \ref{FOF_Finite.p}
(\href{https://raw.githubusercontent.com/GeoffsPapers/InterpretationFormat/master/Examples/FOF_Saturation.s}{Online})}
\label{FOF_Saturation.s}
\begin{small}
\begin{verbatim}
%------------------------------------------------------------------------------
cnf(c_0_15,interpretation-herbrand,
    ( created_equal(X1,X2) | ~ human(X1) | ~ human(X2) ) ).

cnf(c_0_16,interpretation-herbrand, ~ human(grade_of(X1)) ).

cnf(c_0_17,interpretation-herbrand, grade_of(esk3_0) = a ).

cnf(c_0_18,interpretation-herbrand, grade_of(john) = f ).

cnf(c_0_20,interpretation-herbrand,
    ( esk1_0 != esk2_0 | ~ human(esk1_0) | ~ human(esk2_0) ) ).

cnf(c_0_21,interpretation-herbrand,
    ( created_equal(esk1_0,esk2_0) | esk1_0 = esk2_0 ) ).

cnf(c_0_22,interpretation-herbrand, ( human(esk2_0) | esk1_0 = esk2_0 ) ).

cnf(c_0_23,interpretation-herbrand, ( human(esk1_0) | esk1_0 = esk2_0 ) ).

cnf(c_0_24,interpretation-herbrand, ~ human(a) ).

cnf(c_0_25,interpretation-herbrand, ~ human(f) ).

cnf(c_0_26,interpretation-herbrand, a != f ).

cnf(c_0_27,interpretation-herbrand, human(esk3_0) ).

cnf(c_0_28,interpretation-herbrand, human(john) ).
%------------------------------------------------------------------------------
\end{verbatim}
\end{small}

\newpage
\subsection{FOF Verification Problem for \ref{FOF_Finite.p} and \ref{FOF_Formulae.s}
(\href{https://raw.githubusercontent.com/GeoffsPapers/InterpretationFormat/master/Examples/FOF_Formulae.s.p}{Online})}
\label{FOF_Formulae.s.p}
\begin{small}
\begin{verbatim}
%------------------------------------------------------------------------------
fof(created_equal,axiom,
    ! [X0,X1] : ( created_equal(X0,X1) <=> $true ) ).

fof(human,axiom,
    ! [X0] : ( human(X0) <=> ( X0 != "f" & X0 != "a" ) ) ).

fof(john,axiom,
    ! [X0] : ( X0 = john <=> X0 = "john" ) ).

fof(grade_of,axiom,
    ! [X0,X1] :
      ( X0 = grade_of(X1)
    <=> ( ( X0 = "f" & X1 != "gotA" )
        | ( X0 = "a" & X1 = "gotA" ) ) ) ).

fof(f,axiom,
    ! [X0] : ( X0 = f <=> X0 = "f" ) ).

fof(a,axiom,
    ! [X0] : ( X0 = a <=> X0 = "a" ) ).

fof(prove_formulae,conjecture,
    ( ! [H1,H2] :
        ( ( human(H1)
          & human(H2) )
       => created_equal(H1,H2) )
    & human(john)
    & grade_of(john) = f
    & ? [H] :
        ( human(H)
        & grade_of(H) = a )
    & a != f
    & ! [G] : ~ human(grade_of(G))
    & ~ ! [H1,H2] :
          ( ( human(H1)
            & human(H2)
            & created_equal(H1,H2) )
        <=> H1 = H2 ) ) ).
%------------------------------------------------------------------------------
\end{verbatim}
\end{small}

\newpage
\subsection{FOF Verification Problem for \ref{FOF_Finite.p} and \ref{FOF_Saturation.s}
(\href{https://raw.githubusercontent.com/GeoffsPapers/InterpretationFormat/master/Examples/FOF_Saturation.s.p}{Online})}
\label{FOF_Saturation.s.p}
\begin{small}
\begin{verbatim}
%------------------------------------------------------------------------------
cnf(c_0_15,axiom, ( created_equal(X1,X2) | ~ human(X1) | ~ human(X2) ) ).

cnf(c_0_16,axiom, ~ human(grade_of(X1)) ).

cnf(c_0_17,axiom, grade_of(esk3_0) = a ).

cnf(c_0_18,axiom, grade_of(john) = f ).

cnf(c_0_20,axiom, ( esk1_0 != esk2_0 | ~ human(esk1_0) | ~ human(esk2_0) ) ).

cnf(c_0_21,axiom, ( created_equal(esk1_0,esk2_0) | esk1_0 = esk2_0 ) ).

cnf(c_0_22,axiom, ( human(esk2_0) | esk1_0 = esk2_0 ) ).

cnf(c_0_23,axiom, ( human(esk1_0) | esk1_0 = esk2_0 ) ).

cnf(c_0_24,axiom, ~ human(a) ).

cnf(c_0_25,axiom, ~ human(f) ).

cnf(c_0_26,axiom, a != f ).

cnf(c_0_27,axiom, human(esk3_0) ).

cnf(c_0_28,axiom, human(john) ).

fof(prove_formulae,conjecture,
    ( ! [H1,H2] :
        ( ( human(H1)
          & human(H2) )
       => created_equal(H1,H2) )
    & human(john)
    & grade_of(john) = f
    & ? [H] :
        ( human(H)
        & grade_of(H) = a )
    & a != f
    & ! [G] : ~ human(grade_of(G))
    & ~ ! [H1,H2] :
          ( ( human(H1)
            & human(H2)
            & created_equal(H1,H2) )
        <=> H1 = H2 ) ) ).
%------------------------------------------------------------------------------
\end{verbatim}
\end{small}

\newpage
\section{TF0, Finite Interpretations}
\label{TF0Finite}

\subsection{TF0 Problem with a Finite Countermodel
(\href{https://raw.githubusercontent.com/GeoffsPapers/InterpretationFormat/master/Examples/TFF_Finite.p}{Online})}
\label{TFF_Finite.p}
\begin{small}
\begin{verbatim}
%------------------------------------------------------------------------------
tff(human_type,type,      human: $tType ).
tff(cat_type,type,        cat: $tType ).
tff(jon_decl,type,        jon: human ).
tff(garfield_decl,type,   garfield: cat ).
tff(arlene_decl,type,     arlene: cat ).
tff(nermal_decl,type,     nermal: cat ).
tff(loves_decl,type,      loves: cat > cat ).
tff(owns_decl,type,       owns: ( human * cat ) > $o ).

tff(only_jon,axiom, ! [H: human] : H = jon ).

tff(only_garfield_and_arlene_and_nermal,axiom,
    ! [C: cat] : 
      ( C = garfield | C = arlene | C = nermal ) ).

tff(distinct_cats,axiom,
    ( garfield != arlene & arlene != nermal 
    & nermal != garfield ) ).

tff(jon_owns_garfield_not_arlene,axiom,
    ( owns(jon,garfield) & ~ owns(jon,arlene) ) ).

tff(all_cats_love_garfield,axiom,
    ! [C: cat] : ( loves(C) = garfield ) ).

tff(jon_owns_garfields_lovers,conjecture,
    ! [C: cat] :
      ( ( loves(C) = garfield & C != arlene ) 
       => owns(jon,C) ) ).
%------------------------------------------------------------------------------
\end{verbatim}
\end{small}

\newpage
\subsection{TF0 Finite Model for \ref{TFF_Finite.p}, Reusing Problem Types
(\href{https://raw.githubusercontent.com/GeoffsPapers/InterpretationFormat/master/Examples/TFF_Finite.s}{Online})}
\label{TFF_Finite.s}
\begin{small}
\begin{verbatim}
%------------------------------------------------------------------------------
tff(human_type,type,      human: $tType ).
tff(cat_type,type,        cat: $tType ).
tff(jon_decl,type,        jon: human ).
tff(garfield_decl,type,   garfield: cat ).
tff(arlene_decl,type,     arlene: cat ).
tff(nermal_decl,type,     nermal: cat ).
tff(loves_decl,type,      loves: cat > cat ).
tff(owns_decl,type,       owns: ( human * cat ) > $o ).

%----Types of the domain elements
tff(d_jon_decl,type,      d_jon: human ).
tff(d_garfield_decl,type, d_garfield: cat ).
tff(d_arlene_decl,type,   d_arlene: cat ).
tff(d_nermal_decl,type,   d_nermal: cat ).

tff(garfield,interpretation,
%----The human domain elements are {d_jon}
    ( ( ! [DH: human] : ( DH = d_jon )
%----The cat domain elements are {d_garfield,d_arlene,d_nermal}
      & ! [DC: cat]: 
          ( DC = d_garfield | DC = d_arlene | DC = d_nermal )
      & $distinct(d_garfield,d_arlene,d_nermal) )
%----Interpret functions 
    & ( jon = d_jon
      & garfield = d_garfield
      & arlene = d_arlene
      & nermal = d_nermal
      & loves(d_garfield) = d_garfield
      & loves(d_arlene) = d_garfield
      & loves(d_nermal) = d_garfield )
%----Interpret predicates
    & ( owns(d_jon,d_garfield)
      & ~ owns(d_jon,d_arlene)
      & ~ owns(d_jon,d_nermal) ) ) ).
%------------------------------------------------------------------------------
\end{verbatim}
\end{small}

\newpage
\subsection{TF0 Verification Problem for \ref{TFF_Finite.p} and \ref{TFF_Finite.s}
(\href{https://raw.githubusercontent.com/GeoffsPapers/InterpretationFormat/master/Examples/TFF_Finite.s.p}{Online})}
\label{TFF_Finite.s.p}
\begin{small}
\begin{verbatim}
%------------------------------------------------------------------------------
tff(human_type,type,      human: $tType ).
tff(cat_type,type,        cat: $tType ).
tff(jon_decl,type,        jon: human ).
tff(garfield_decl,type,   garfield: cat ).
tff(arlene_decl,type,     arlene: cat ).
tff(nermal_decl,type,     nermal: cat ).
tff(loves_decl,type,      loves: cat > cat ).
tff(owns_decl,type,       owns: ( human * cat ) > $o ).

tff(d_jon_decl,type,      d_jon: human ).
tff(d_garfield_decl,type, d_garfield: cat ).
tff(d_arlene_decl,type,   d_arlene: cat ).
tff(d_nermal_decl,type,   d_nermal: cat ).

tff(garfield,axiom,
%----The human domain elements are {d_jon}
    ( ( ! [DH: human] : ( DH = d_jon )
%----The cat domain elements are {d_garfield,d_arlene,d_nermal}
      & ! [DC: cat]: 
          ( DC = d_garfield | DC = d_arlene | DC = d_nermal )
      & $distinct(d_garfield,d_arlene,d_nermal) )
%----Interpret terms via the type-promoted domain
    & ( jon = d_jon
      & garfield = d_garfield
      & arlene = d_arlene
      & nermal = d_nermal
      & loves(d_garfield) = d_garfield
      & loves(d_arlene) = d_garfield
      & loves(d_nermal) = d_garfield )
%----Interpret atoms as true or false
    & ( owns(d_jon,d_garfield)
      & ~ owns(d_jon,d_arlene)
      & ~ owns(d_jon,d_nermal) ) ) ).

tff(verify,conjecture,
    ( ! [H: human] : H = jon
    & ! [C: cat] : ( C = garfield | C = arlene | C = nermal )
    & ( garfield != arlene & garfield != nermal & arlene != nermal )
    & ( owns(jon,garfield) & ~ owns(jon,arlene) )
    & ! [C: cat] : ( loves(C) = garfield )
    & ~ ! [C: cat] :
          ( ( loves(C) = garfield & C != arlene)
         => owns(jon,C) ) ) ).
%------------------------------------------------------------------------------
\end{verbatim}
\end{small}

\newpage
\subsection{TF0 Finite Model for \ref{TFF_Finite.p}, Separate Domain Types
(\href{https://raw.githubusercontent.com/GeoffsPapers/InterpretationFormat/master/Examples/TFF_Finite_SeparateDomains.s}{Online})}
\label{TFF_Finite_SeparateDomains.s}
\begin{small}
\begin{verbatim}
%------------------------------------------------------------------------------
tff(human_type,type,      human: $tType ).
tff(cat_type,type,        cat: $tType ).
tff(jon_decl,type,        jon: human ).
tff(garfield_decl,type,   garfield: cat ).
tff(arlene_decl,type,     arlene: cat ).
tff(nermal_decl,type,     nermal: cat ).
tff(loves_decl,type,      loves: cat > cat ).
tff(owns_decl,type,       owns: ( human * cat ) > $o ).

%----Types of the domains
tff(d_human_type,type,    d_human: $tType ).
tff(d_cat_type,type,      d_cat: $tType ).
%----Types of the promotion functions
tff(d2human_decl,type,    d2human: d_human > human ).
tff(d2cat_decl,type,      d2cat: d_cat > cat ).
%----Types of the domain elements
tff(d_jon_decl,type,      d_jon: d_human ).
tff(d_garfield_decl,type, d_garfield: d_cat ).
tff(d_arlene_decl,type,   d_arlene: d_cat ).
tff(d_nermal_decl,type,   d_nermal: d_cat ).

tff(garfield,interpretation,
%----The domain for human is d_human
    ( ( ! [H: human] : ? [DH: d_human] : H = d2human(DH)
%----The d_human elements are {d_jon}
      & ! [DH: d_human] : ( DH = d_jon )
%----The type-promoter is a bijection
      & ! [DH1: d_human,DH2: d_human] :
          ( d2human(DH1) = d2human(DH2) => DH1 = DH2 )
%----The domain for cat is d_cat
      & ! [C: cat] : ? [DC: d_cat] : C = d2cat(DC)
%----The d_cat elements are {d_garfield,d_arlene,d_nermal}
      & ! [DC: d_cat]: 
          ( DC = d_garfield | DC = d_arlene | DC = d_nermal )
      & $distinct(d_garfield,d_arlene,d_nermal)
%----The type-promoter is a bijection
      & ! [DC1: d_cat,DC2: d_cat] : 
          ( d2cat(DC1) = d2cat(DC2) => DC1 = DC2 ) )
%----Interpret terms via the type-promoted domain
    & ( jon = d2human(d_jon)
      & garfield = d2cat(d_garfield)
      & arlene = d2cat(d_arlene)
      & nermal = d2cat(d_nermal)
      & loves(d2cat(d_garfield)) = d2cat(d_garfield)
      & loves(d2cat(d_arlene)) = d2cat(d_garfield)
      & loves(d2cat(d_nermal)) = d2cat(d_garfield) )
%----Interpret atoms as true or false
    & ( owns(d2human(d_jon),d2cat(d_garfield))
      & ~ owns(d2human(d_jon),d2cat(d_arlene))
      & ~ owns(d2human(d_jon),d2cat(d_nermal)) ) ) ).
%------------------------------------------------------------------------------
\end{verbatim}
\end{small}

\newpage
\subsection{TF0 Verification Problem for \ref{TFF_Finite.p} and \ref{TFF_Finite_SeparateDomains.s}
(\href{https://raw.githubusercontent.com/GeoffsPapers/InterpretationFormat/master/Examples/TFF_Finite_SeparateDomains.s.p}{Online})}
\label{TFF_Finite_SeparateDomains.s.p}
\begin{small}
\begin{verbatim}
%------------------------------------------------------------------------------
tff(human_type,type,      human: $tType ).
tff(cat_type,type,        cat: $tType ).
tff(jon_decl,type,        jon: human ).
tff(garfield_decl,type,   garfield: cat ).
tff(arlene_decl,type,     arlene: cat ).
tff(nermal_decl,type,     nermal: cat ).
tff(loves_decl,type,      loves: cat > cat ).
tff(owns_decl,type,       owns: ( human * cat ) > $o ).

%----Types of the domains
tff(d_human_type,type,    d_human: $tType ).
tff(d_cat_type,type,      d_cat: $tType ).
%----Types of the promotion functions
tff(d2human_decl,type,    d2human: d_human > human ).
tff(d2cat_decl,type,      d2cat: d_cat > cat ).
%----Types of the domain elements
tff(d_jon_decl,type,      d_jon: d_human ).
tff(d_garfield_decl,type, d_garfield: d_cat ).
tff(d_arlene_decl,type,   d_arlene: d_cat ).
tff(d_nermal_decl,type,   d_nermal: d_cat ).

tff(garfield,axiom,
%----The domain for human is d_human
    ( ( ! [H: human] : ? [DH: d_human] : H = d2human(DH)
%----The d_human elements are {d_jon}
      & ! [DH: d_human] : ( DH = d_jon )
%----The type-promoter is a bijection
      & ! [DH1: d_human,DH2: d_human] :
          ( d2human(DH1) = d2human(DH2) => DH1 = DH2 )
%----The domain for cat is d_cat
      & ! [C: cat] : ? [DC: d_cat] : C = d2cat(DC)
%----The d_cat elements are {d_garfield,d_arlene,d_nermal}
      & ! [DC: d_cat]:
          ( DC = d_garfield | DC = d_arlene | DC = d_nermal )
      & $distinct(d_garfield,d_arlene,d_nermal)
%----The type-promoter is a bijection
      & ! [DC1: d_cat,DC2: d_cat] :
          ( d2cat(DC1) = d2cat(DC2) => DC1 = DC2 ) )
%----Interpret terms via the type-promoted domain
    & ( jon = d2human(d_jon)
      & garfield = d2cat(d_garfield)
      & arlene = d2cat(d_arlene)
      & nermal = d2cat(d_nermal)
      & loves(d2cat(d_garfield)) = d2cat(d_garfield)
      & loves(d2cat(d_arlene)) = d2cat(d_garfield)
      & loves(d2cat(d_nermal)) = d2cat(d_garfield) )
%----Interpret atoms as true or false
    & ( owns(d2human(d_jon),d2cat(d_garfield))
      & ~ owns(d2human(d_jon),d2cat(d_arlene))
      & ~ owns(d2human(d_jon),d2cat(d_nermal)) ) ) ).

tff(verify,conjecture,
    ( ! [H: human] : H = jon
    & ! [C: cat] : ( C = garfield | C = arlene | C = nermal )
    & ( garfield != arlene & garfield != nermal & arlene != nermal )
    & ( owns(jon,garfield) & ~ owns(jon,arlene) )
    & ! [C: cat] : ( loves(C) = garfield )
    & ~ ! [C: cat] :
          ( ( loves(C) = garfield & C != arlene)
         => owns(jon,C) ) ) ).
%------------------------------------------------------------------------------
\end{verbatim}
\end{small}

\newpage
\subsection{TF0 Finite Model for \ref{TFF_Finite.p}, Medium Grained
(\href{https://raw.githubusercontent.com/GeoffsPapers/InterpretationFormat/master/Examples/TFF_Finite_Medium.s}{Online})}
\label{TFF_Finite_Medium.s}
\begin{small}
\begin{verbatim}
%------------------------------------------------------------------------------
tff(human_type,type,      human: $tType ).
tff(cat_type,type,        cat: $tType ).
tff(jon_decl,type,        jon: human ).
tff(garfield_decl,type,   garfield: cat ).
tff(arlene_decl,type,     arlene: cat ).
tff(nermal_decl,type,     nermal: cat ).
tff(loves_decl,type,      loves: cat > cat ).
tff(owns_decl,type,       owns: ( human * cat ) > $o ).

tff(d_human_type,type,    d_human: $tType ).
tff(d_cat_type,type,      d_cat: $tType ).
tff(d2human_decl,type,    d2human: d_human > human ).
tff(d2cat_decl,type,      d2cat: d_cat > cat ).
tff(d_jon_decl,type,      d_jon: d_human ).
tff(d_garfield_decl,type, d_garfield: d_cat ).
tff(d_arlene_decl,type,   d_arlene: d_cat ).
tff(d_nermal_decl,type,   d_nermal: d_cat ).

tff(garfield_domains,interpretation-domains,
    ( ! [H: human] : ? [DH: d_human] : H = d2human(DH)
    & ! [DH: d_human] : ( DH = d_jon )
    & ! [DH1: d_human,DH2: d_human] :
        ( d2human(DH1) = d2human(DH2) => DH1 = DH2 )
    & ! [C: cat] : ? [DC: d_cat] : C = d2cat(DC)
    & ! [DC: d_cat]: 
        ( DC = d_garfield | DC = d_arlene | DC = d_nermal )
    & $distinct(d_garfield,d_arlene,d_nermal)
    & ! [DC1: d_cat,DC2: d_cat] : 
        ( d2cat(DC1) = d2cat(DC2) => DC1 = DC2 ) ) ).

tff(garfield_mappings,interpretation-mappings,
    ( ( jon = d2human(d_jon)
      & garfield = d2cat(d_garfield)
      & arlene = d2cat(d_arlene)
      & nermal = d2cat(d_nermal)
      & loves(d2cat(d_garfield)) = d2cat(d_garfield)
      & loves(d2cat(d_arlene)) = d2cat(d_garfield)
      & loves(d2cat(d_nermal)) = d2cat(d_garfield) )
    & ( owns(d2human(d_jon),d2cat(d_garfield))
      & ~ owns(d2human(d_jon),d2cat(d_arlene))
      & ~ owns(d2human(d_jon),d2cat(d_nermal)) ) ) ).
%------------------------------------------------------------------------------
\end{verbatim}
\end{small}

\newpage
\subsection{TF0 Finite Model for \ref{TFF_Finite.p}, Fine Grained
(\href{https://raw.githubusercontent.com/GeoffsPapers/InterpretationFormat/master/Examples/TFF_Finite_Fine.s}{Online})}
\label{TFF_Finite_Fine.s}
\begin{small}
\begin{verbatim}
%------------------------------------------------------------------------------
tff(human_type,type,      human: $tType ).
tff(cat_type,type,        cat: $tType ).
tff(jon_decl,type,        jon: human ).
tff(garfield_decl,type,   garfield: cat ).
tff(arlene_decl,type,     arlene: cat ).
tff(nermal_decl,type,     nermal: cat ).
tff(loves_decl,type,      loves: cat > cat ).
tff(owns_decl,type,       owns: ( human * cat ) > $o ).

tff(d_human_type,type,    d_human: $tType ).
tff(d_cat_type,type,      d_cat: $tType ).
tff(d2human_decl,type,    d2human: d_human > human ).
tff(d2cat_decl,type,      d2cat: d_cat > cat ).
tff(d_jon_decl,type,      d_jon: d_human ).
tff(d_garfield_decl,type, d_garfield: d_cat ).
tff(d_arlene_decl,type,   d_arlene: d_cat ).
tff(d_nermal_decl,type,   d_nermal: d_cat ).

tff(garfield_domain_human,interpretation-domain(human,d_human),
    ( ! [H: human] : ? [DH: d_human] : H = d2human(DH)
    & ! [DH: d_human] : ( DH = d_jon )
    & ! [DH1: d_human,DH2: d_human] :
        ( d2human(DH1) = d2human(DH2) => DH1 = DH2 ) ) ).

tff(garfield_domain_cat,interpretation-domain(cat,d_cat),
    ( ! [C: cat] : ? [DC: d_cat] : C = d2cat(DC)
    & ! [DC: d_cat]: 
        ( DC = d_garfield | DC = d_arlene | DC = d_nermal )
    & $distinct(d_garfield,d_arlene,d_nermal)
    & ! [DC1: d_cat,DC2: d_cat] : 
        ( d2cat(DC1) = d2cat(DC2) => DC1 = DC2 ) ) ).

tff(garfield_mapping_jon,interpretation-mapping(jon,d_human),
    jon = d2human(d_jon) ).

tff(garfield_mapping_garfield,interpretation-mapping(garfied,d_cat),
    garfield = d2cat(d_garfield) ).

tff(garfield_mapping_arlene,interpretation-mapping(arlene,d_cat),
    arlene = d2cat(d_arlene) ).

tff(garfield_mapping_nermal,interpretation-mapping(nermal,d_cat),
    nermal = d2cat(d_nermal) ).

tff(garfield_mapping_loves,interpretation-mapping(loves,d_cat),
    ( loves(d2cat(d_garfield)) = d2cat(d_garfield)
    & loves(d2cat(d_arlene)) = d2cat(d_garfield)
    & loves(d2cat(d_nermal)) = d2cat(d_garfield) ) ).

tff(garfield_mapping_owns,interpretation-mapping(owns,$o),
    ( owns(d2human(d_jon),d2cat(d_garfield))
    & ~ owns(d2human(d_jon),d2cat(d_arlene))
    & ~ owns(d2human(d_jon),d2cat(d_nermal)) ) ).
%------------------------------------------------------------------------------
\end{verbatim}
\end{small}

\newpage
\subsection{TF0 Finite Model for \ref{TFF_Finite.p}, Compacted
(\href{https://raw.githubusercontent.com/GeoffsPapers/InterpretationFormat/master/Examples/TFF_Finite_Compact.s}{Online})}
\label{TFF_Finite_Compact.s}
\begin{small}
\begin{verbatim}
%------------------------------------------------------------------------------
tff(human_type,type,      human: $tType ).
tff(cat_type,type,        cat: $tType ).
tff(jon_decl,type,        jon: human ).
tff(garfield_decl,type,   garfield: cat ).
tff(arlene_decl,type,     arlene: cat ).
tff(nermal_decl,type,     nermal: cat ).
tff(loves_decl,type,      loves: cat > cat ).
tff(owns_decl,type,       owns: ( human * cat ) > $o ).

tff(d_human_type,type,    d_human: $tType ).
tff(d_cat_type,type,      d_cat: $tType ).
tff(d2human_decl,type,    d2human: d_human > human ).
tff(d2cat_decl,type,      d2cat: d_cat > cat ).
tff(d_jon_decl,type,      d_jon: d_human ).
tff(d_garfield_decl,type, d_garfield: d_cat ).
tff(d_arlene_decl,type,   d_arlene: d_cat ).
tff(d_nermal_decl,type,   d_nermal: d_cat ).

tff(garfield_domain_human,interpretation-domains(human,d_human),
    ( ! [H: human] : ? [DH: d_human] : H = d2human(DH)
    & ! [DH: d_human] : ( DH = d_jon )
    & ! [DH1: d_human,DH2: d_human] :
        ( d2human(DH1) = d2human(DH2) => DH1 = DH2 ) ) ).

tff(garfield_domain_cat,interpretation-domains(cat,d_cat),
    ( ! [C: cat] : ? [DC: d_cat] : C = d2cat(DC)
    & ! [DC: d_cat]: 
        ( DC = d_garfield | DC = d_arlene | DC = d_nermal )
    & $distinct(d_garfield,d_arlene,d_nermal)
    & ! [DC1: d_cat,DC2: d_cat] : 
        ( d2cat(DC1) = d2cat(DC2) => DC1 = DC2 ) ) ).

tff(garfield_mapping_jon,interpretation-mappings(jon,d_human),
    jon = d2human(d_jon) ).

tff(garfield_mapping_cats,interpretation-mappings,
    ( garfield = d2cat(d_garfield)
    & arlene = d2cat(d_arlene)
    & nermal = d2cat(d_nermal) ) ).

tff(garfield_mapping_loves,interpretation-mappings(loves,d_cat),
    ! [DC: d_cat] : ( loves(d2cat(DC)) = d2cat(d_garfield) ) ).

tff(garfield_mapping_owns,interpretation-mappings(owns,$o),
    ( owns(d2human(d_jon),d2cat(d_garfield))
    & ~ owns(d2human(d_jon),d2cat(d_arlene))
    & ~ owns(d2human(d_jon),d2cat(d_nermal)) ) ).
%------------------------------------------------------------------------------
\end{verbatim}
\end{small}

\newpage
\section{TF0, Infinite Interpretations}
\label{TF0Infinite}

\subsection{TF0 Axioms with an Infinite Model
(\href{https://raw.githubusercontent.com/GeoffsPapers/InterpretationFormat/master/Examples/TFF_Infinite.p}{Online})}
\label{TFF_Infinite.p}
\begin{small}
\begin{verbatim}
%------------------------------------------------------------------------------
tff(person_type,type,        person: $tType ).
tff(bob_decl,type,           bob: person ).
tff(child_of_decl,type,      child_of: person > person ).
tff(is_descendant_decl,type, is_descendant: (person * person) > $o ).

tff(descendents_different,axiom,
    ! [A: person,D: person] : 
      ( is_descendant(A,D) => ( A != D ) ) ).

tff(descendent_transitive,axiom,
    ! [A: person,C: person,G: person] :
      ( ( is_descendant(A,C) & is_descendant(C,G) ) 
     => is_descendant(A,G) ) ).

tff(child_is_descendant,axiom,
    ! [P: person] : is_descendant(P,child_of(P)) ).

tff(all_have_child,axiom,
    ! [P: person] : ? [C: person] : C = child_of(P) ).
%------------------------------------------------------------------------------
\end{verbatim}
\end{small}

\newpage
\subsection{TF0 Infinite Model for \ref{TFF_Infinite.p}, Term Domain
(\href{https://raw.githubusercontent.com/GeoffsPapers/InterpretationFormat/master/Examples/TFF_Peano.s}{Online})}
\label{TFF_Peano.s}
\begin{small}
\begin{verbatim}
%------------------------------------------------------------------------------
tff(person_type,type,         person: $tType ).
tff(bob_decl,type,            bob: person ).
tff(child_of_decl,type,       child_of: person > person ).
tff(is_descendant_decl,type,  is_descendant: ( person * person ) > $o ).

tff(peano_type,type,          peano: $tType).
tff(zero_decl,type,           zero: peano ).
tff(s_decl,type,              s: peano > peano ).
tff(peano2person_decl,type,   peano2person: peano > person ).
tff(peano_less_decl,type,     peano_less: ( peano * peano ) > $o ).

tff(people,interpretation,
%----Domain for type person is the Peano numbers
    ( ! [P: person] : ? [I: peano] : P = peano2person(I)
    & ! [I: peano] : ( I = zero | ? [P: peano] : I = s(P) )
%----The type promoter is a bijection (injective and surjective)
    & ! [I1: peano,I2: peano] :
        ( peano2person(I1) = peano2person(I2) => I1 = I2 )
%----Relationships between Peano numbers
    & ! [I1: peano,I2: peano,I3: peano] :
        ( peano_less(I1,s(I1))
        & ( ( peano_less(I1,I2) & peano_less(I2,I3) )
         => peano_less(I1,I3) )
        & ( peano_less(I1,I2)
         => I1 != I2 ) )
%----Mapping people to Peano numbers
    & bob = peano2person(zero)
    & ! [I: peano] :
        child_of(peano2person(I)) = peano2person(s(I))
%----Interpretation of descendancy
    & ! [A: peano,D: peano] :
        ( is_descendant(peano2person(A),peano2person(D))
      <=> peano_less(A,D) ) ) ).
%------------------------------------------------------------------------------
\end{verbatim}
\end{small}

\newpage
\subsection{TF0 Infinite Model for \ref{TFF_Infinite.p}, Integer Domain
(\href{https://raw.githubusercontent.com/GeoffsPapers/InterpretationFormat/master/Examples/TFF_Integer.s}{Online})}
\label{TFF_Integer.s}
\begin{small}
\begin{verbatim}
%------------------------------------------------------------------------------
tff(person_type,type,        person: $tType ).
tff(bob_decl,type,           bob: person ).
tff(child_of_decl,type,      child_of: person > person ).
tff(is_descendant_decl,type, is_descendant: ( person * person ) > $o ).

tff(int2person_decl,type,    int2person: $int > person ).

tff(people,interpretation,
%----Domain for type person is the integers
    ( ! [P: person] : ? [I: $int] : int2person(I) = P
%----The type promoter is a bijection (injective and surjective)
    & ! [I1: $int,I2: $int] : 
        ( int2person(I1) = int2person(I2) => I1 = I2 )
%----Mapping people to integers. Note that Bob's ancestors will be interpreted 
%----as negative integers.
    & bob = int2person(0)
    & ! [I: $int] : child_of(int2person(I)) = int2person($sum(I,1))
%----Interpretation of descendancy
    & ! [A: $int,D: $int] : 
        ( is_descendant(int2person(A),int2person(D)) <=> $less(A,D) ) ) ).
%------------------------------------------------------------------------------
\end{verbatim}
\end{small}

\newpage
\subsection{TF0 Verification Problem for \ref{TFF_Infinite.p} and \ref{TFF_Peano.s}
(\href{https://raw.githubusercontent.com/GeoffsPapers/InterpretationFormat/master/Examples/TFF_Peano.s.p}{Online})}
\label{TFF_Peano.s.p}
\begin{small}
\begin{verbatim}
%------------------------------------------------------------------------------
tff(person_type,type,         person: $tType ).
tff(bob_decl,type,            bob: person ).
tff(child_of_decl,type,       child_of: person > person ).
tff(is_descendant_decl,type,  is_descendant: ( person * person ) > $o ).

tff(peano_type,type,          peano: $tType).
tff(zero_decl,type,           zero: peano ).
tff(s_decl,type,              s: peano > peano ).
tff(peano2person_decl,type,peano2person: peano > person ).
tff(peano_less_decl,type,     peano_less: ( peano * peano ) > $o ).

tff(people,axiom,
    ( ( ! [P: person] : ? [I: peano] : ( P = peano2person(I) )
      & ! [I: peano] : ( I = zero | ? [P: peano] : I = s(P) )
      & ! [I1: peano,I2: peano] :
          ( peano2person(I1) = peano2person(I2) => I1 = I2 ) 
      & ! [I1: peano,I2: peano,I3: peano] :
          ( peano_less(I1,s(I1))
          & ( ( peano_less(I1,I2)
              & peano_less(I2,I3) )
           => peano_less(I1,I3) )
          & ( peano_less(I1,I2)
           => I1 != I2 ) ) )
    & ( bob = peano2person(zero)
      & ! [I: peano] :
          child_of(peano2person(I)) = peano2person(s(I)) )
    & ( ! [A: peano,D: peano] :
          ( is_descendant(peano2person(A),peano2person(D)) 
        <=> peano_less(A,D) ) ) ) ).

tff(prove_formulae,conjecture,
    ( ! [A: person,D: person] : ( is_descendant(A,D) => A != D )
    & ! [A: person,C: person,G: person] :
        ( ( is_descendant(A,C) 
          & is_descendant(C,G) )
       => is_descendant(A,G) )
    & ! [P: person] : is_descendant(P,child_of(P))
    & ! [P: person] : ? [C: person] : C = child_of(P) ) ).
%------------------------------------------------------------------------------
\end{verbatim}
\end{small}

\newpage
\subsection{TF0 Verification Problem for \ref{TFF_Infinite.p} and \ref{TFF_Integer.s}
(\href{https://raw.githubusercontent.com/GeoffsPapers/InterpretationFormat/master/Examples/TFF_Integer.s.p}{Online})}
\label{TFF_Integer.s.p}
\begin{small}
\begin{verbatim}
%------------------------------------------------------------------------------
tff(person_type,type,        person: $tType ).
tff(bob_decl,type,           bob: person ).
tff(child_of_decl,type,      child_of: person > person ).
tff(is_descendant_decl,type, is_descendant: ( person * person ) > $o ).

tff(int2person_decl,type,    int2person: $int > person ).

tff(people,axiom,
    ( ( ! [P: person] : ? [I: $int] : int2person(I) = P
      & ! [I1: $int,I2: $int] : 
          ( int2person(I1) = int2person(I2) => I1 = I2 ) )
    & ( bob = int2person(0)
      & ! [I: $int] : child_of(int2person(I)) = int2person($sum(I,1)) )
    & ! [A: $int,D: $int] : 
        ( is_descendant(int2person(A),int2person(D)) <=> $less(A,D) ) ) ).

tff(prove_formulae,conjecture,
    ( ! [A: person,D: person] : ( is_descendant(A,D) => A != D )
    & ! [A: person,C: person,G: person] :
        ( ( is_descendant(A,C) & is_descendant(C,G) ) => is_descendant(A,G) )
    & ! [P: person] : is_descendant(P,child_of(P))
    & ! [P: person] : ? [C: person] : C = child_of(P) ) ).
%------------------------------------------------------------------------------
\end{verbatim}
\end{small}

\newpage
\section{TH0, Finite Interpretations}
\label{TH0Finite}

\subsection{TH0 Problem with a Finite Countermodel
(\href{https://raw.githubusercontent.com/GeoffsPapers/InterpretationFormat/master/Examples/THF_Finite.p}{Online})}
\label{THF_Finite.p}
\begin{small}
\begin{verbatim}
%------------------------------------------------------------------------------
thf(beverage_type,type,   beverage: $tType ).
thf(syrup_type,type,      syrup: $tType ).
thf(coffee_decl,type,     coffee: beverage ).
thf(vanilla_decl,type,    vanilla: syrup ).
thf(mix_decl,type,        mix: ( syrup > beverage ) > beverage ).
thf(heat_decl,type,       heat: beverage > beverage ).
thf(heated_mix_decl,type, heated_mix: ( syrup > beverage ) > beverage ).
thf(hot_decl,type,        hot: beverage > $o ).

thf(heated_mix,axiom,
    ( heated_mix
    = ( ^ [F: syrup > beverage] : ( heat @ ( mix @ F ) ) ) ) ).

thf(hot_mixture,axiom,
    ! [F: syrup > beverage] : ( hot @ ( heated_mix @ F ) ) ).

thf(heated_coffee_mix,axiom,
    ! [F: syrup > beverage] : ( ( heated_mix @ F ) = coffee ) ).

thf(hot_coffee,conjecture,
    ? [Mixture: syrup > beverage] :
      ~ ? [S: syrup] :
          ( ( ( Mixture @ S ) = coffee )
          & ( hot @ ( Mixture @ S ) ) ) ).
%------------------------------------------------------------------------------
\end{verbatim}
\end{small}

\newpage
\subsection{TH0 Finite Model for \ref{THF_Finite.p}, Coarse Grained
(\href{https://raw.githubusercontent.com/GeoffsPapers/InterpretationFormat/master/Examples/THF_Finite.s}{Online})}
\label{THF_Finite.s}
\begin{small}
\begin{verbatim}
%------------------------------------------------------------------------------
thf(beverage_type,type,   beverage: $tType ).
thf(syrup_type,type,      syrup: $tType ).
thf(coffee_decl,type,     coffee: beverage ).
thf(vanilla_decl,type,    vanilla: syrup ).
thf(mix_decl,type,        mix: ( syrup > beverage ) > beverage ).
thf(heat_decl,type,       heat: beverage > beverage ).
thf(heated_mix_decl,type, heated_mix: ( syrup > beverage ) > beverage ).
thf(hot_decl,type,        hot: beverage > $o ).

thf(d_beverage_type,type, d_beverage: $tType ).
thf(d_syrup_type,type,    d_syrup: $tType ).
thf(d2beverage_decl,type, d2beverage: d_beverage > beverage ).
thf(d2syrup_decl,type,    d2syrup: d_syrup > syrup ).
thf(d_coffee_decl,type,   d_coffee: d_beverage ).
thf(d_vanilla_decl,type,  d_vanilla: d_syrup ).

thf(hot_coffee,interpretation,
    ( ! [B: beverage] : ? [DB: d_beverage] : ( B = ( d2beverage @ DB ) )
    & ! [DB: d_beverage] : ( DB = d_coffee )
    & ! [DB1: d_beverage,DB2: d_beverage] :
        ( ( ( d2beverage @ DB1 ) = ( d2beverage @ DB2 ) )
       => ( DB1 = DB2 ) )
    & ! [S: syrup] : ? [DS: d_syrup] : ( S = ( d2syrup @ DS ) )
    & ! [DS: d_syrup] : ( DS = d_vanilla )
    & ! [DS1: d_syrup,DS2: d_syrup] :
        ( ( ( d2syrup @ DS1 ) = ( d2syrup @ DS2 ) )
       => ( DS1 = DS2 ) )
    & ( coffee = ( d2beverage @ d_coffee ) )
    & ( vanilla = ( d2syrup @ d_vanilla ) )
    & ( ( heat @ ( d2beverage @ d_coffee ) ) = d_coffee )
    & ( mix = ( ^ [F: syrup > beverage] : ( d2beverage @ d_coffee ) ) )
    & ( heated_mix
      = ( ^ [F: syrup > beverage] : ( d2beverage @ d_coffee ) ) )
    & ( hot @ ( d2beverage @ d_coffee ) ) ) ).
%------------------------------------------------------------------------------
\end{verbatim}
\end{small}

\newpage
\subsection{TH0 Verification Problem for \ref{THF_Finite.p} and \ref{THF_Finite.s}
(\href{https://raw.githubusercontent.com/GeoffsPapers/InterpretationFormat/master/Examples/THF_Finite.s.p}{Online})}
\label{THF_Finite.s.p}
\begin{small}
\begin{verbatim}
%------------------------------------------------------------------------------
thf(beverage_type,type,   beverage: $tType ).
thf(syrup_type,type,      syrup: $tType ).
thf(coffee_decl,type,     coffee: beverage ).
thf(vanilla_decl,type,    vanilla: syrup ).
thf(mix_decl,type,        mix: ( syrup > beverage ) > beverage ).
thf(heat_decl,type,       heat: beverage > beverage ).
thf(heated_mix_decl,type, heated_mix: ( syrup > beverage ) > beverage ).
thf(hot_decl,type,        hot: beverage > $o ).

thf(d_beverage_type,type, d_beverage: $tType ).
thf(d_syrup_type,type,    d_syrup: $tType ).
thf(d2beverage_decl,type, d2beverage: d_beverage > beverage ).
thf(d2syrup_decl,type,    d2syrup: d_syrup > syrup ).
thf(d_coffee_decl,type,   d_coffee: d_beverage ).
thf(d_vanilla_decl,type,  d_vanilla: d_syrup ).

thf(model,axiom,
    ( ! [B: beverage] : ? [DB: d_beverage] : ( B = ( d2beverage @ DB ) )
    & ! [DB: d_beverage] : ( DB = d_coffee )
    & ! [DB1: d_beverage,DB2: d_beverage] :
        ( ( ( d2beverage @ DB1 ) = ( d2beverage @ DB2 ) )
       => ( DB1 = DB2 ) )
    & ! [S: syrup] : ? [DS: d_syrup] : ( S = ( d2syrup @ DS ) )
    & ! [DS: d_syrup] : ( DS = d_vanilla )
    & ! [DS1: d_syrup,DS2: d_syrup] :
        ( ( ( d2syrup @ DS1 ) = ( d2syrup @ DS2 ) )
       => ( DS1 = DS2 ) )
    & ( coffee = ( d2beverage @ d_coffee ) )
    & ( vanilla = ( d2syrup @ d_vanilla ) )
    & ( ( heat @ ( d2beverage @ d_coffee ) ) = d_coffee )
    & ( mix = ( ^ [F: syrup > beverage] : ( d2beverage @ d_coffee ) ) )
    & ( heated_mix
      = ( ^ [F: syrup > beverage] : ( d2beverage @ d_coffee ) ) )
    & ( hot @ ( d2beverage @ d_coffee ) ) ) ).

thf(verify,conjecture,
    ( ( heated_mix
      = ( ^ [F: syrup > beverage] : ( heat @ ( mix @ F ) ) ) )
    & ! [F: syrup > beverage] : ( hot @ ( heated_mix @ F ) )
    & ! [F: syrup > beverage] : ( ( heated_mix @ F ) = coffee )
    &  ~ ? [Mixture: syrup > beverage] :
           ~ ? [S: syrup] :
               ( ( ( Mixture @ S ) = coffee )
               & ( hot @ ( Mixture @ S ) ) ) ) ).
%------------------------------------------------------------------------------
\end{verbatim}
\end{small}

\newpage
\subsection{TH0 Finite Model for \ref{THF_Finite.p}, Mixed Grained
(\href{https://raw.githubusercontent.com/GeoffsPapers/InterpretationFormat/master/Examples/THF_Finite_Medium.s}{Online})}
\label{THF_Finite_Medium.s}
\begin{small}
\begin{verbatim}
%------------------------------------------------------------------------------
thf(beverage_type,type,   beverage: $tType ).
thf(syrup_type,type,      syrup: $tType ).
thf(coffee_decl,type,     coffee: beverage ).
thf(vanilla_decl,type,    vanilla: syrup ).
thf(mix_decl,type,        mix: ( beverage > syrup ) > beverage ).
thf(heat_decl,type,       heat: beverage > beverage ).
thf(heated_mix_decl,type, heated_mix: ( beverage > syrup ) > beverage ).
thf(hot_decl,type,        hot: beverage > $o ).

thf(d_beverage_type,type, d_beverage: $tType ).
thf(d_syrup_type,type,    d_syrup: $tType ).
thf(d2beverage_decl,type, d2beverage: d_beverage > beverage ).
thf(d2syrup_decl,type,    d2syrup: d_syrup > syrup ).
thf(d_coffee_decl,type,   d_coffee: d_beverage ).
thf(d_vanilla_decl,type,  d_vanilla: d_syrup ).

thf(hot_coffee_beverage,interpretation-domains(beverage,d_beverage),
    ( ! [B: beverage] : ? [DB: d_beverage] : ( B = ( d2beverage @ DB ) )
    & ! [DB: d_beverage] : ( DB = d_coffee )
    & ! [DB1: d_beverage,DB2: d_beverage] :
        ( ( ( d2beverage @ DB1 ) = ( d2beverage @ DB2 ) )
       => ( DB1 = DB2 ) ) ) ).

thf(hot_coffee_syrup,interpretation-domains(syrup,d_syrup),
    ( ! [S: syrup] : ? [DS: d_syrup] : ( S = ( d2syrup @ DS ) )
    & ! [DS: d_syrup] : ( DS = d_vanilla )
    & ! [DS1: d_syrup,DS2: d_syrup] :
        ( ( ( d2syrup @ DS1 ) = ( d2syrup @ DS2 ) )
       => ( DS1 = DS2 ) ) ) ).

thf(hot_coffee_mappings,interpretation-mappings,
    ( ( coffee = ( d2beverage @ d_coffee ) )
    & ( vanilla = ( d2syrup @ d_vanilla ) )
    & ( ( heat @ ( d2beverage @ d_coffee ) ) = d_coffee )
    & ( mix = ( ^ [F: beverage > syrup] : ( d2beverage @ d_coffee ) ) )
    & ( heated_mix
      = ( ^ [F: beverage > syrup] : ( d2beverage @ d_coffee ) ) )
    & ( hot @ ( d2beverage @ d_coffee ) ) ) ).
%------------------------------------------------------------------------------
\end{verbatim}
\end{small}

\newpage
\section{NXF, Finite Worlds with Finite Interpretations}
\label{NXF}

\subsection{NXF Problem with Global Axioms, \\ with a Finite-Finite Countermodel
(\href{https://raw.githubusercontent.com/GeoffsPapers/InterpretationFormat/master/Examples/NXF_Finite-Finite-Global.p}{Online})}
\label{NXF_Finite-Finite-Global.p}
\begin{small}
\begin{verbatim}
%------------------------------------------------------------------------------
tff(semantics,logic,
    $alethic_modal ==
      [ $domains == $constant,
        $designation == $rigid,
        $terms == $local,
        $modalities == $modal_system_K ] ).

tff(child_type,type,    child: $tType ).
tff(adult_type,type,    adult: $tType ).
tff(agatha_decl,type,   agatha: adult ).
tff(charly_decl,type,   charly: child ).
tff(quiet_decl,type,    quiet: child > $o ).
tff(sleepy_decl,type,   sleepy: adult > $o ).
tff(peaceful_decl,type, peaceful: child > $o ).
tff(serves_decl,type,   serves: adult > child ).
tff(rains_decl,type,    rains: $o ).

tff(a1,axiom,
    ! [C: child] :
      ~ ( ~quiet(C) & ? [A: adult] : sleepy(A) ) ).

tff(a2,axiom,
    ( ( rains & ? [C: child] : quiet(C) )
   => ! [C: child] : ~peaceful(C) ) ).

tff(a3,axiom,
    ( peaceful(charly)
    | ( ~quiet(charly) & quiet(serves(agatha)) ) ) ).

tff(a4,axiom,
    ~quiet(charly) => ! [C: child] : ~quiet(C) ).

tff(a5,axiom,
    {$necessary} @ ( peaceful(charly) => rains ) ).

tff(c,conjecture,
    {$possible} @ (~ rains) ).
%------------------------------------------------------------------------------
\end{verbatim}
\end{small}

\newpage
\subsection{TXF Finite-Finite Model for \ref{NXF_Finite-Finite-Global.p}
(\href{https://raw.githubusercontent.com/GeoffsPapers/InterpretationFormat/master/Examples/NXF_Finite-Finite-Global.s}{Online})}
\label{NXF_Finite-Finite-Global.s}
\begin{small}
\begin{verbatim}
%------------------------------------------------------------------------------
tff(semantics,logic,
    $modal ==
      [ $domains == $constant,
        $designation == $rigid,
        $terms == $local,
        $modalities == $modal_system_K ] ).

tff(child_type,type,    child: $tType ).
tff(adult_type,type,    adult: $tType ).
tff(agatha_decl,type,   agatha: adult ).
tff(charly_decl,type,   charly: child ).
tff(quiet_decl,type,    quiet: child > $o ).
tff(sleepy_decl,type,   sleepy: adult > $o ).
tff(peaceful_decl,type, peaceful: child > $o ).
tff(serves_decl,type,   serves: adult > child ).
tff(rains_decl,type,    rains: $o ).

tff(child_d_decl,type, child_d: $tType).
tff(adult_d_decl,type, adult_d: $tType).
tff(child_1_decl,type, child_1: child_d).
tff(adult_1_decl,type, adult_1: adult_d).
tff(d2child_decl,type, d2child: child_d > child).
tff(d2adult_decl,type, d2adult: adult_d > adult).

tff(w1_decl,type,       w1: $world ).
tff(w2_decl,type,       w2: $world ).
tff(w3_decl,type,       w3: $world ).

tff(people_worlds,interpretation,
    ( ( ! [W: $world] :
          ( W = w1 | W = w2 | W = w3 )
      & $distinct(w1,w2,w3)
      & $local_world = w1
      & $accessible_world(w1,w1) & $accessible_world(w2,w2)
      & $accessible_world(w1,w2) & $accessible_world(w2,w3)
      & $accessible_world(w3,w1)
      & ~ $accessible_world(w1,w3) & ~ $accessible_world(w2,w1)
      & ~ $accessible_world(w3,w2) & ~ $accessible_world(w3,w3) )
    & $in_world(w1,
        ( ! [C: child] : ? [CD: child_d] : C = d2child(CD)
        & ! [CD: child_d] : CD = child_1
        & ? [CD: child_d] : CD = child_1
        & ! [CD1: child_d,CD2: child_d] :
            ( d2child(CD1) = d2child(CD2) => CD1 = CD2 )
        & ! [A: adult] : ? [AD: adult_d] : A = d2adult(AD)
        & ! [AD: adult_d] : AD = adult_1
        & ? [AD: adult_d] : AD = adult_1
        & ! [AD1: adult_d,AD2: adult_d] :
            ( d2adult(AD1) = d2adult(AD2) => AD1 = AD2 )
        & charly = d2child(child_1)
        & agatha = d2adult(adult_1)
        & serves(d2adult(adult_1)) = d2child(child_1)
        & ~ quiet(d2child(child_1))
        & ~ sleepy(d2adult(adult_1))
        & peaceful(d2child(child_1))
        & rains ))
    & $in_world(w2,
        ( ! [C: child] : ? [CD: child_d] : C = d2child(CD)
        & ! [CD: child_d] : CD = child_1
        & ? [CD: child_d] : CD = child_1
        & ! [CD1: child_d,CD2: child_d] :
            ( d2child(CD1) = d2child(CD2) => CD1 = CD2 )
        & ! [A: adult] : ? [AD: adult_d] : A = d2adult(AD)
        & ! [AD: adult_d] : AD = adult_1
        & ? [AD: adult_d] : AD = adult_1
        & ! [AD1: adult_d,AD2: adult_d] :
            ( d2adult(AD1) = d2adult(AD2) => AD1 = AD2 )
        & charly = d2child(child_1)
        & agatha = d2adult(adult_1)
        & serves(d2adult(adult_1)) = d2child(child_1)
        & ~ quiet(d2child(child_1))
        & ~ sleepy(d2adult(adult_1))
        & peaceful(d2child(child_1))
        & rains ))
    & $in_world(w3,
        ( ! [C: child] : ? [CD: child_d] : C = d2child(CD)
        & ! [CD: child_d] : CD = child_1
        & ? [CD: child_d] : CD = child_1
        & ! [CD1: child_d,CD2: child_d] :
            ( d2child(CD1) = d2child(CD2) => CD1 = CD2 )
        & ! [A: adult] : ? [AD: adult_d] : A = d2adult(AD)
        & ! [AD: adult_d] : AD = adult_1
        & ? [AD: adult_d] : AD = adult_1
        & ! [AD1: adult_d,AD2: adult_d] :
            ( d2adult(AD1) = d2adult(AD2) => AD1 = AD2 )
        & charly = d2child(child_1)
        & agatha = d2adult(adult_1)
        & serves(d2adult(adult_1)) = d2child(child_1)
        & ~ quiet(d2child(child_1))
        & ~ sleepy(d2adult(adult_1))
        & peaceful(d2child(child_1))
        & rains )) ) ).
%------------------------------------------------------------------------------
\end{verbatim}
\end{small}

\newpage
\subsection{NXF Verification Problem for \ref{NXF_Finite-Finite-Global.p} and 
\ref{NXF_Finite-Finite-Global.s}
(\href{https://raw.githubusercontent.com/GeoffsPapers/InterpretationFormat/master/Examples/NXF_Finite-Finite-Global.s.p}{Online})}
\label{NXF_Finite-Finite-Global.s.p}
\begin{small}
\begin{verbatim}
%------------------------------------------------------------------------------
tff(the_logic,logic,$$fomlModel).

tff(child_type,type,    child: $tType ).
tff(adult_type,type,    adult: $tType ).
tff(agatha_decl,type,   agatha: adult ).
tff(charly_decl,type,   charly: child ).
tff(quiet_decl,type,    quiet: child > $o ).
tff(sleepy_decl,type,   sleepy: adult > $o ).
tff(peaceful_decl,type, peaceful: child > $o ).
tff(serves_decl,type,   serves: adult > child ).
tff(rains_decl,type,    rains: $o ).

tff(child_d_decl,type, child_d: $tType).
tff(adult_d_decl,type, adult_d: $tType).
tff(child_1_decl,type, child_1: child_d).
tff(adult_1_decl,type, adult_1: adult_d).
tff(d2child_decl,type, d2child: child_d > child).
tff(d2adult_decl,type, d2adult: adult_d > adult).

tff(w1_decl,type,       w1: $world ).
tff(w2_decl,type,       w2: $world ).
tff(w3_decl,type,       w3: $world ).

tff(people_worlds,interpretation,
    ( ( ! [W: $world] :
          ( W = w1 | W = w2 | W = w3 )
      & $distinct(w1,w2,w3)
      & $local_world = w1
      & $accessible_world(w1,w1) & $accessible_world(w2,w2)
      & $accessible_world(w1,w2) & $accessible_world(w2,w3)
      & $accessible_world(w3,w1)
      & ~ $accessible_world(w1,w3) & ~ $accessible_world(w2,w1)
      & ~ $accessible_world(w3,w2) & ~ $accessible_world(w3,w3) )
    & $in_world(w1,
        ( ! [C: child] : ? [CD: child_d] : C = d2child(CD)
        & ! [CD: child_d] : CD = child_1
        & ? [CD: child_d] : CD = child_1
        & ! [CD1: child_d,CD2: child_d] :
            ( d2child(CD1) = d2child(CD2) => CD1 = CD2 )
        & ! [A: adult] : ? [AD: adult_d] : A = d2adult(AD)
        & ! [AD: adult_d] : AD = adult_1
        & ? [AD: adult_d] : AD = adult_1
        & ! [AD1: adult_d,AD2: adult_d] :
            ( d2adult(AD1) = d2adult(AD2) => AD1 = AD2 )
        & charly = d2child(child_1)
        & agatha = d2adult(adult_1)
        & serves(d2adult(adult_1)) = d2child(child_1)
        & ~ quiet(d2child(child_1))
        & ~ sleepy(d2adult(adult_1))
        & peaceful(d2child(child_1))
        & rains ))
    & $in_world(w2,
        ( ! [C: child] : ? [CD: child_d] : C = d2child(CD)
        & ! [CD: child_d] : CD = child_1
        & ? [CD: child_d] : CD = child_1
        & ! [CD1: child_d,CD2: child_d] :
            ( d2child(CD1) = d2child(CD2) => CD1 = CD2 )
        & ! [A: adult] : ? [AD: adult_d] : A = d2adult(AD)
        & ! [AD: adult_d] : AD = adult_1
        & ? [AD: adult_d] : AD = adult_1
        & ! [AD1: adult_d,AD2: adult_d] :
            ( d2adult(AD1) = d2adult(AD2) => AD1 = AD2 )
        & charly = d2child(child_1)
        & agatha = d2adult(adult_1)
        & serves(d2adult(adult_1)) = d2child(child_1)
        & ~ quiet(d2child(child_1))
        & ~ sleepy(d2adult(adult_1))
        & peaceful(d2child(child_1))
        & rains ))
    & $in_world(w3,
        ( ! [C: child] : ? [CD: child_d] : C = d2child(CD)
        & ! [CD: child_d] : CD = child_1
        & ? [CD: child_d] : CD = child_1
        & ! [CD1: child_d,CD2: child_d] :
            ( d2child(CD1) = d2child(CD2) => CD1 = CD2 )
        & ! [A: adult] : ? [AD: adult_d] : A = d2adult(AD)
        & ! [AD: adult_d] : AD = adult_1
        & ? [AD: adult_d] : AD = adult_1
        & ! [AD1: adult_d,AD2: adult_d] :
            ( d2adult(AD1) = d2adult(AD2) => AD1 = AD2 )
        & charly = d2child(child_1)
        & agatha = d2adult(adult_1)
        & serves(d2adult(adult_1)) = d2child(child_1)
        & ~ quiet(d2child(child_1))
        & ~ sleepy(d2adult(adult_1))
        & peaceful(d2child(child_1))
        & rains )) ) ).

tff(a1,conjecture-global,
    ! [C: child] :
      ~ ( ~ quiet(C)
        & ? [A: adult] : sleepy(A) ) ).

tff(a2,conjecture-global,
    ( ( rains
      & ? [C: child] : quiet(C) )
   => ! [C: child] : ~ peaceful(C) ) ).

tff(a3,conjecture-global,
    ( peaceful(charly)
    | ( ~ quiet(charly)
      & quiet(serves(agatha)) ) ) ).

tff(a4,conjecture-global,
    ( ~ quiet(charly)
   => ! [C: child] : ~ quiet(C) ) ).

tff(a5,conjecture-global,
    ( {$necessary}
    @ ( ( peaceful(charly)
       => rains ) ) ) ).

tff(c,conjecture-local,
    ~ ( {$possible} @ (~ rains) ) ).
%------------------------------------------------------------------------------
\end{verbatim}
\end{small}


\newpage
\subsection{NXF Problem with Global and Local Axioms, \\
            with a Finite-Finite Countermodel
(\href{https://raw.githubusercontent.com/GeoffsPapers/InterpretationFormat/master/Examples/NXF_Finite-Finite-Local.p}{Online})}
\label{NXF_Finite-Finite-Local.p}
\begin{small}
\begin{verbatim}
%------------------------------------------------------------------------------
tff(semantics,logic,
    $alethic_modal == 
      [ $domains == $constant,
        $designation == $rigid,
        $terms == $local,
        $modalities == $modal_system_M ] ).

tff(fruit_type,type,   fruit: $tType).
tff(apple_decl,type,   apple: fruit).
tff(banana_decl,type,  banana: fruit).
tff(healthy_decl,type, healthy: fruit > $o).
tff(rotten_decl,type,  rotten: fruit > $o).

tff(apple_not_banana,axiom,
    apple != banana ).

tff(necessary_healthy_fruit_everywhere,axiom,
    ! [F: fruit] : ( {$necessary} @ (healthy(F)) ) ).

tff(fruit_possibly_not_rotten,axiom,
    ! [F: fruit] : ( {$possible} @ (~ rotten(F)) ) ).

tff(rotten_banana_here,axiom-local,
    rotten(banana) ).

tff(not_true,conjecture,
    ( {$necessary} @ 
      (( healthy(apple)
       & ~ rotten(banana) )) ) ).
%------------------------------------------------------------------------------
\end{verbatim}
\end{small}

\newpage
\subsection{TXF Finite-Finite Model for \ref{NXF_Finite-Finite-Local.p}
(\href{https://raw.githubusercontent.com/GeoffsPapers/InterpretationFormat/master/Examples/NXF_Finite-Finite-Local.s}{Online})}
\label{NXF_Finite-Finite-Local.s}
\begin{small}
\verbatiminput{Examples/NXF_Finite-Finite-Local.s}
\end{small}

\newpage
\subsection{NXF Verification Problem for \ref{NXF_Finite-Finite-Local.p} and
\ref{NXF_Finite-Finite-Local.s}
(\href{https://raw.githubusercontent.com/GeoffsPapers/InterpretationFormat/master/Examples/NXF_Finite-Finite-Local.s.p}{Online})}
\label{NXF_Finite-Finite-Local.s.p}
\begin{small}
\verbatiminput{Examples/NXF_Finite-Finite-Local.s.p}
\end{small}


\newpage
\subsection{TXF Finite-Finite Model for \ref{NXF_Finite-Finite-Global.p}, Medium Grained \\
(\href{https://raw.githubusercontent.com/GeoffsPapers/InterpretationFormat/master/Examples/NXF_Finite-Finite-Global_Medium.s}{Online})}
\label{NXF_Finite-Finite-Global_Medium.s}
\begin{small}
\verbatiminput{Examples/NXF_Finite-Finite-Global_Medium.s}
\end{small}

\newpage
\subsection{TXF Finite-Finite Model for \ref{NXF_Finite-Finite-Global.p}, Fine Grained
(\href{https://raw.githubusercontent.com/GeoffsPapers/InterpretationFormat/master/Examples/NXF_Finite-Finite-Global_Fine.s}{Online})}
\label{NXF_Finite-Finite-Global_Fine.s}
\begin{small}
\verbatiminput{Examples/NXF_Finite-Finite-Global_Fine.s}
\end{small}

\newpage
\subsection{TXF Finite-Finite Model for \ref{NXF_Finite-Finite-Global.p}, Compacted
(\href{https://raw.githubusercontent.com/GeoffsPapers/InterpretationFormat/master/Examples/NXF_Finite-Finite-Global_Compact.s}{Online})}
\label{NXF_Finite-Finite-Global_Compact.s}
\begin{small}
\verbatiminput{Examples/NXF_Finite-Finite-Global_Compact.s}
\end{small}

\end{document}